\documentclass{iopart}

\pdfoutput=1

\usepackage{graphicx,epsfig,sidecap,wasysym}
\usepackage{bm}% bold math

\def\be{\begin{equation}}
\def\ee{\end{equation}}

\def\bea{\begin{eqnarray}}
\def\eea{\end{eqnarray}}

\def\Tr{{\rm Tr}}

\def\e{\epsilon}

\def\a{\alpha}
\def\b{\beta}
\def\la{\lambda}

\def\HH{\mathcal H}
\def\pl{\prod\limits}
\def\sul{\sum\limits}
\def\e{\epsilon}

\def\nm{\newmoon}
\def\fm{\fullmoon}

\newcommand{\bra}[1]{\langle\,#1\,|}
\newcommand{\ket}[1]{|\,#1\,\rangle}

\begin{document}

\title[Entanglement entropy  of excited states]
{Entanglement entropy of excited states}

\author{Vincenzo Alba$^1$, Maurizio Fagotti$^{2}$, and Pasquale Calabrese$^2$}
\address{$^1$Scuola Normale Superiore and INFN, Pisa, Italy.\\
         $^2$Dipartimento di Fisica dell'Universit\`a di Pisa and INFN,
             Pisa, Italy.}

\date{\today}

\begin{abstract}

We study the entanglement entropy of a block of contiguous spins in excited states 
of spin chains. 
We consider the XY model in a transverse field and the XXZ Heisenberg spin-chain. 
For the latter, we developed a numerical application of algebraic Bethe Ansatz. 
We find two main classes of states with logarithmic and extensive behavior in the 
dimension of the block, characterized by the properties of excitations of the state. 
This behavior can be related to the locality properties of the Hamiltonian having 
a given state as ground state. 
We also provide several details of the finite size scaling.

\end{abstract}

\maketitle

\section{Introduction}

The study of the entanglement in the ground-states of
extended quantum systems became a major enterprise in recent times,
mainly because of its ability in detecting the scaling behavior in
proximity of quantum critical points (see e.g. 
Refs. \cite{af-rev,e-rev,ccd-sp}  as reviews). 
The most studied measure of entanglement is surely the entanglement entropy 
$S_A$,  defined as follows. 
Let $\rho$ be the density matrix of a system, which we take to be in the 
pure quantum state $|\Psi\rangle$, $\rho=|\Psi\rangle\langle\Psi|$.
Let the Hilbert space be written as a direct product $\HH=\HH_A\otimes\HH_B$. 
$A$'s reduced density matrix is $\rho_A=\Tr_B\, \rho$.
The entanglement entropy is the corresponding von Neumann entropy
\be
S_A=-\Tr\, \rho_A \log \rho_A\,,
\label{Sdef}
\ee
and analogously for $S_B$. 
When $\rho$ corresponds to a pure quantum state $S_A=S_B$. 

The entanglement entropy is one of the best indicators of the 
critical properties of an extended quantum system when $A$ and $B$ are 
a spatial bipartition of the system. 
Well-known and fundamental examples are critical one-dimensional systems in the 
case when $A$ is an interval of length $\ell$ in a system of length $N$
with periodic boundary conditions. In this case, the entanglement entropy follows the scaling \cite{cc-04,cc-rev}
\be
S_A= \frac{c}{3} \log\left(\frac{N}{\pi}
\sin\frac{\pi \ell}{N}\right)+ c_1'
\stackrel{N\to\infty}{\longrightarrow}
\frac{c}3 \log\ell +c_1'\,,
\label{SAlog}
\ee
where $c$ is the central charge of the underlying
conformal field theory and $c'_1$ a non-universal constant 
(the behavior for $N\to\infty$ is known from Refs. \cite{Holzhey,Vidal}).
Away from the critical point, $S_A$ saturates to a constant value
\cite{Vidal} proportional to the logarithm of the correlation length
\cite{cc-04}.
This scaling allows to locate the position (where $S_A$ diverges by 
increasing $\ell$) and a main feature (the value of the central charge $c$) 
of quantum critical points displaying conformal invariance.
The entanglement entropy of disjoint intervals gives also information 
about other universal features of the conformal fixed point related to the 
full operator content of the theory \cite{multi}.

Conversely, only little attention has been devoted to the entanglement 
properties of excited states (with the exception of few manuscripts 
\cite{ds-3,m-05,as-08,m-09}), although it is a very natural problem. 
Here we consider two topical spin-chains \cite{lr-rev} to address this issue.
We first consider the XY model in a transverse magnetic field. 
We employ the well-known mapping of the model to free fermions to reduce the 
calculation of the entanglement entropy to that of the eigenvalues of a
Toeplitz matrix on the lines of the ground-state case \cite{Vidal,p-dm,p-04,pe-rev,jk-04,ijk-05,ij-08}.
In the present computation, the properties of the excitations above the 
ground-state will strongly affect the form of the reduced density matrix and of
the entanglement entropy. 
Then, to consider a truly strongly interacting quantum model, we address the 
same problem for the XXZ chain, always remaining in the realm of integrable 
systems. In fact, this model is exactly solvable 
by means of Bethe Ansatz \cite{BetheZP71,GaudinBOOK}. 
This provides a classification of all eigenstates and their energies, but no 
information about dynamical properties. To overcome this limit, we take advantage 
of recent progresses in the algebraic Bethe Ansatz \cite{kmt-99,kmt-00} 
that provides all elements of the reduced density matrix 
as a (huge) sum of determinants 
whose entries are functions of the Bethe rapidities. However, in this approach an 
inhomogeneous coupling must be considered and the homogeneous limit (in which we are 
interested) is recovered in a cumbersome manner.

In the study of the entanglement properties of excited states, a first subtle 
point is the choice of the basis of the Hilbert space. 
In fact, while the ground-state of a local Hamiltonian is usually unique 
(or with a finite small degeneracy, when some symmetry is not spontaneously broken),
the excited states can be highly degenerate. Thus, any linear combination of
them is still an eigenstate. In principle the entanglement properties
can vary a lot with the basis. However, we will show that some of our findings 
are general features of all linear combinations of the same class of excited states.
This is not only an academic subtlety, because the exact studies one can 
perform are limited to integrable models, for which it is well-known 
that the degeneracy is large. Oppositely, any small integrability breaking 
term will remove these degenerations and one could wonder whether 
the specific properties found are only features of integrable models. 

The quantification of the entanglement in excited states can have consequences 
in the understanding of the quantum out-of-equilibrium physics and in particular 
of the dynamical problems known as {\it quantum quenches}. 
In fact, it has been argued that the post-quench state is a time-dependent 
superposition of eigenstates that in the thermodynamic limit have the same
energy \cite{s08}. 
It is known that for a {\it global} quench, the entanglement entropy first increases 
linearly with the time and then saturates to a values proportional to the 
length of the block $\ell$ \cite{cc-05,QE,fc-08}. 
We will indeed find a full class of excited states having an extensive 
entanglement entropy and those could be the relevant ones for quench problems. 
Oppositely in {\it local} quantum quenches the asymptotic state displays a logarithmic
entanglement entropy \cite{LQE} and a different class of states should be relevant.
We also mention that some of the features we find have similarities 
with what obtained in some non-equilibrium steady states \cite{ness}.

The manuscript is organized as follows. In the next section \ref{sec2} we study the XY model and 
we find two main classes of excited states, corresponding the extensive and logarithmic
behavior of the entanglement entropy.  In Sec. \ref{sec3} we consider the XXZ model 
and the algebraic Bethe ansatz approach. 
We find that the states that have a logarithmic behavior in the XX limit conserve this 
property with the same prefactor of the logarithm and with a constant term slightly 
depending on $\Delta$.
Finally in Sec. \ref{concl} we summarize our main results and discuss problems 
deserving further investigation.

\section{The XY model in a transverse magnetic field}
\label{sec2}

We start our analysis by considering the XY spin chain of length $N$ with 
periodic boundary conditions, whose Hamiltonian is given by
\begin{equation}\label{eq:Hamiltonian}
H_{XY}=-\sum_{l=1}^N \left[
J \left(\frac{1+\gamma}{4}\sigma_l^x\sigma_{l+1}^x
+\frac{1-\gamma}{4}\sigma_l^y\sigma_{l+1}^y\right)+\frac{h}{2}\sigma_l^z
\right]\, ,
\end{equation}
where \(\sigma_l^{\alpha}\) are the Pauli matrices at the site $l$.
$h$ is the transverse magnetic field and $\gamma$ the anisotropy parameter.
For $\gamma=1$ the Hamiltonian reduces to the Ising model, while for 
$\gamma=0$ to the XX model.
The diagonalization of this Hamiltonian is a standard textbook exercise. 
First a Jordan-Wigner transformation
\be
c_l = \left( \prod_{m<l}\sigma_m^z \right) \frac{\sigma_l^x -
 i\sigma_l^y}{2}\,, 
\qquad
c_l^{\dagger} = \left( \prod_{m<l}\sigma_m^z \right) \frac{\sigma_l^x +
 i\sigma_l^y}{2}\,,
\ee
maps the model into a quadratic spinless free-fermion Hamiltonian (i.e. 
with anticommutation relations
$\{c_l^{\dagger},c_m\}=\delta_{lm}$ , $\{c_l,c_m\} =0$).
After Fourier transforming in  momentum space 
$c_k =\sum_l c_l e^{-i\frac{2\pi}{N}kl}/\sqrt{N}$, 
the so-called Bogoliubov transformation 
\be
b^{\dagger}_k =u_k \, c^{\dagger}_k + i v_k \, c_{-k}\,, \qquad 
b_k = u_k \, c_k - i v_k \, c^{\dagger}_{-k},
\label{bogo}
\ee
makes the Hamiltonian diagonal
\be
H=\sum_{k=\frac{1-N}{2}}^{\frac{N-1}{2}}\varepsilon_k 
\left(b^\dag_k b_k^{\phantom{\dag}}-\frac12\right)\,,
\ee
where we considered $N$ to be odd. 
We ignored a boundary term that gives a vanishing contribution in the thermodynamic limit.
Here we introduced the Bogoliubov variables
$u_k = \cos{\theta_k/2},~ v_k = \sin{\theta_k/2}$ and angle
\be
\tan\theta_k=\frac{J\gamma\sin\varphi_k}{J\cos\varphi_k-h}\,,
\qquad {\rm with} \;\; \varphi_k=\frac{2\pi k}{N}\,,
\label{bogangle}
\ee
giving single-particle eigenvalues
\be
\varepsilon_k= \sqrt{(h-J\cos\varphi_k)^2+J^2 \gamma^2 \sin^2\varphi_k}\,.
\ee
From this dispersion relation, it is evident that the model is 
critical (gapless) for $\gamma=0$ and $|h|<|J|$ (XX universality class)
and for $h=\pm J$ and any $\gamma\neq0$ (Ising universality class). 
%In the following we fix $J=1$ that sets the energy scale.  

The exact diagonalization of the model gives not only the ground-state 
properties but a complete classification of all the eigenstates and in 
particular their energy.
In the basis of free fermions, the excited states are classified
according to the occupation numbers of the single-particle basis 
(that is the basis of Slater determinants). A generic eigenstate
can be written as 
\be\label{EX}\fl
|E_x\rangle \equiv\prod_{k\in E_x}b^\dag_k|0\rangle\,,
\qquad {\rm with\; energy}\;\; E_{E_x}=\frac12 \left(\sum_{k\in E_x}\varepsilon_k-
\sum_{k\notin E_x}\varepsilon_k\right)\,,
\label{ES}
\ee 
where $E_x$ is the set of occupied momenta. To give a simple pictorial 
representation of these states, we indicate with up-arrows 
the occupied single-particle levels (excited quasiparticles) 
and with down-arrows the empty ones, with the first arrow corresponding to momentum 
$\varphi_k=-\pi$. 
When a set of $n$ consecutive momenta are occupied (empty), we simply replace
the up (down) string with $\uparrow^n$ ($\downarrow^n$).
For example, the ground state is 
$|\downarrow\dots \downarrow\rangle=|\downarrow^N\rangle$. 
Counting all the possible arrow orientations, it is obvious that this 
graphical representation generates all the $2^N$ eigenstates of the chain.
Notice that these arrows have nothing to do with the state of the spin in real 
space (the real space configuration is an highly entangled superposition).

When calculating the entanglement entropy, three different length scales 
enter in the computation: the size of the chain $N$, the length of the block 
$\ell$ and the number of excited quasiparticles that is encoded in the size 
$|E_x|$ of the set $E_x$. 
General results can be obtained in the thermodynamic limit $N\to\infty$ and
when $\ell\gg1$ (in finite size, this limit describes the regime 
$N\gg\ell\gg1$).
In this limit, it is obvious that if only a small number of quasiparticle 
levels are populated  (i.e. $|E_x|\ll N$), 
the corrections to the ground-state correlation matrix 
can be generally treated perturbatively and in a first approximation the 
excited quasiparticles contribute independently to the entanglement, giving 
rise to a negligible contribution in $1/N$. 
Thus, in the thermodynamic limit, all entanglement properties of these
states are equivalent to those of the ground-state, but this does not prevent 
from interesting and maybe calculable finite-size behavior.
We have been informed of some unpublished work by M. Ibanez and G. Sierra \cite{is}
studying the entanglement entropy of these low-lying excited states obtaining 
a finite-size scaling different from Eq. (\ref{SAlog}). 
Here instead we are interested in those states that are macroscopically 
different from the ground-state and that will have an 
entanglement entropy that in the thermodynamic limit could differ strongly from 
Eq.  (\ref{SAlog}).

In order to work directly in the thermodynamic limit, we need a proper
description of excited states. This is rather straightforward. 
In fact, when $N\to\infty$ the possible values of $k$ are all the integer 
numbers and the reduced momentum $\varphi_k$ becomes a continuous variable 
$\varphi$ living in the interval $\varphi\in]-\pi,\pi[$. 
We are here interested in the case with $|E_x|\sim N$ (that can be seen as 
an ``highly excited state'', even if it is not the energy that matters).
Thus in all the formulas involving sums over populated energy levels,
we substitute sums with integrals by using as distribution a proper defined 
regularized {\it characteristic function} of the set $E_x$ that we will indicate
as $m(\varphi)$. The function $(1+m(\varphi))/2$ represents the average occupation of 
levels in an infinitesimal shell around the momentum $\varphi_k=2\pi k/N$. 
Let us give several examples to make this limiting procedure clear ($\alpha<1$):
\bea
|\downarrow^N\rangle&&\longrightarrow m(\varphi)=-1\,,\nonumber\\
|\downarrow^{N/2}\uparrow^{\alpha N/2}\downarrow^{N(1-\alpha)/2}\rangle&&
\longrightarrow m(\varphi)=
\cases{1\,,& if $0\leq\varphi<\pi \alpha$\,,\cr
       -1\,& otherwise\,,}\nonumber\\
|\downarrow^{\alpha N/2}\uparrow^{N(1-\alpha)}\downarrow^{\alpha N/2}\rangle&&
\longrightarrow m(\varphi)=
\cases{1\,,& if $|\varphi|<\pi \alpha$\,,\cr
       -1\,& otherwise\,,}\nonumber\\
|\{\uparrow\downarrow\}^{N/2}\rangle&&\longrightarrow m(\varphi)=0\,,\nonumber\\
|\{\downarrow^2\uparrow\}^{N/3}\rangle&&\longrightarrow m(\varphi)=-1/3\,,
\nonumber\\
|\{\downarrow^2\uparrow\}^{N/6}|\{\uparrow^2\downarrow\}^{N/6}\rangle&&
\longrightarrow m(\varphi)=
\cases{-1/3\,,& if $-\pi<\varphi<0 $\,,\cr
       1/3\,& otherwise\,.}
\label{mdef}
\eea
We only wrote down for simplicity states with a step-wise characteristic 
function, but with little fantasy it is easy to imagine states 
with a smooth one \footnote{If we would be pedantic in defining this limit, 
we can think to 
$(1+m(\varphi))/{2}$ as the convolution of the characteristic function of $E_x$
with a Gaussian of zero mean and standard deviation that must be put to zero 
at the end of any computation. Since in the sum in Eq.
(\ref{eq:Correlations}) there is almost everywhere (everywhere in non-critical regions) 
a regular function of $\varphi$, 
the regularization in the definition of $m(\varphi)$ is perfectly well-defined.}.

\subsection{The reduced density matrix and the entanglement entropy}

It has been shown \cite{Vidal,p-dm} that, despite the non-local character 
of the Jordan-Wigner transformation, the spectrum of the reduced density 
matrix $\rho_A$ of a single interval $A=[0,\ell]$ is the same in the 
spin variables $\sigma_l$ and in the free-fermion ones $c_l$. 
This property makes the XY model the ideal testing-ground 
to understand the behavior of the single-block entanglement for the excited states. 
The eigenvalues of the reduced density matrix 
$\rho_{\ell}$ of a block of $\ell$ adjacent spins for a Slater determinant 
are related to the eigenvalues $\nu_i$ of the correlation matrix 
restricted to the subsystem \cite{Vidal,p-dm}. 
This is easier to see by introducing the Majorana operators 
$A_l^x=c^\dag_l+c_l^{\phantom{\dag}}$ and $A_l^y=i(c_l^{\phantom{\dag}}-c^\dag_l)$
\cite{Vidal}. 
The eigenvalues of $\rho_\ell$ can be labelled with the 
configurations of $\ell$ 
classical spin-variables denoted as $\tau_j=\pm1$  and it holds 
$\lambda_{\{\tau\}}=\prod_{j=1}^\ell(1+\tau_j \nu_j)/{2}$, 
with $i \nu_j$ the eigenvalues of the block Toeplitz  matrix 
\begin{equation}\label{eq:CorrelationMatrix}
\fl \Pi=\left(\begin{array}{ccc}
\Gamma_0 & \cdots &\Gamma_{\ell-1} \\
\vdots&\ddots&\vdots\\
\Gamma_{1-\ell}&\cdots&\Gamma_0
\end{array}\right)\,,\qquad\quad 
\Gamma_l=\Bigl<\left(\begin{array}{c} A_s^x\\A_s^y
\end{array}\right)
\left(\begin{array}{cc} 
A_{s+l}^x&A_{s+l}^y
\end{array}\right)\Bigr>-\mathbf{1}\delta_{l 0}\, .
\end{equation}
The two-by-two matrices $\Gamma_l$ are easily computed observing that the 
generic eigenstate in the Slater-determinant basis (\ref{ES})
is the vacuum of the fermionic operators 
\[
\tilde b^\dag_k, =\cases{ 
b_k, &$k\in E_x$,\cr 
b^\dag_k& otherwise. } 
\]
After simple algebra one obtains
\begin{equation}\label{eq:Correlations}
\Gamma_l^{(E_x)}\!\!\!=\Gamma_l^{(GS)}\!\!\!+\frac{2i}{N}\sum_{k\in E_x}
\left(\begin{array}{cc}
\sin(l \varphi_k)&-\cos(l \varphi_k-\theta_k)\\
\cos(l\varphi_k+\theta_k)&\sin(l\varphi_k)
\end{array}\right)\, ,
\end{equation}
where \(\theta_k\) is the Bogolioubov angle of the transformation that 
diagonalizes the Hamiltonian in Eq. (\ref{bogangle}) and $\Gamma_l^{(GS)}$ 
the corresponding matrix in the ground-state \cite{Vidal, p-dm}.

As explained in the previous subsection, when $|E_x|\sim N$, we can substitute 
in equation (\ref{eq:Correlations}) the sum with an integral 
\be
\frac{1}{N}\sum_{k\in E_x}\rightarrow
\frac{1}{2\pi}\int_{-\pi}^\pi \mathrm{d}\varphi 
\frac{1+m(\varphi)}{2}\qquad \varphi_k\rightarrow \varphi\, ,
\ee
where $(1+m(\varphi))/{2}$ is the regularized characteristic function of 
the set $E_x$ introduced above.
Substituting in Eq. (\ref{eq:Correlations}) this regularization we have
\bea
&&\fl \Gamma_l^{(E_x)}=\frac{1}{2\pi}
\int_{-\pi}^\pi\mathrm{d}\varphi e^{-il\varphi}\Gamma^{(E_x)}_\varphi\,, 
\quad {\rm with}\\
\label{eq:Gamma}
&& \Gamma^{(E_x)}_\varphi=\frac{1}{2}
\left(\begin{array}{cc}
m(-\varphi)-m(\varphi)&-i[m(\varphi)+m(-\varphi)]e^{i\theta}\\
i[m(\varphi)+m(-\varphi)]e^{-i\theta}&m(-\varphi)-m(\varphi)
\end{array}\right)\, .
\eea

The entanglement entropy can be expressed as a complex integration over a
contour $C$ that encircles the segment $[-1,1]$ at the infinitesimal 
distance $\eta$ as in Ref. \cite{jk-04} 
\begin{equation}\label{eq:Scontour}
S_\ell=\lim_{\eta\rightarrow 0^+}\frac{1}{4\pi i}
\oint_C \mathrm{d}\lambda e(1+2\eta,\lambda)\frac{\mathrm d}{\mathrm d \lambda}
\log \det|\lambda \mathbf{1}-\Pi|\, ,
\end{equation}
where 
$$e(x,y)=-\frac{x+y}{2}\log\frac{x+y}{2}-\frac{x-y}{2}\log\frac{x-y}{2}\,.$$
A similar expression is easily written for all R\'enyi entropies for 
general $n$.
Applying the Sz\"ego lemma (see e.g. Ref. \cite{am-74}) to the determinant of the 
block Toeplitz matrix $\lambda \mathbf{1}-\Pi$, we obtain the leading order 
in $\ell$ of the entanglement entropy 
\begin{equation}\label{eq:extensive}
S_\ell= \frac{\ell}{2\pi}\int_{-\pi}^\pi\mathrm{d}\varphi\ H(m(\varphi))+O(\log\ell)\,,
\end{equation}
with $H(x)=e(1,x)$.

This first result is very suggestive: the entanglement 
entropy of a class of excited states in the XY model is extensive, 
in contrast with the logarithmic behavior of the ground state.  
However, every time that  $m(\varphi)^2\neq 1$ only in a 
region of vanishing measure of the domain (as in the ground state) this leading term vanishes, 
and one should go beyond the Sz\"ego lemma to derive the first non-vanishing
order of the entanglement entropy. 
It is important to stress that for this type of ``highly excited states'' 
the leading order of the entanglement entropy is not sensitive of the 
criticality of the ground-state. This does not come unexpected, because we 
are exploring a region of energy that lies extensively above the ground-state.

To describe the (subleading) logarithmic terms in the determinant of a 
Toeplitz matrix, we should use the so-called Fisher-Hartwig 
conjecture \cite{fh}. 
If $m(\varphi)^2= 1$ almost everywhere, $m(\varphi)$ can be re-written 
in the following form, that is particularly useful to apply Fisher-Hartwig
($\varphi\in]-\pi,\pi[$)
\begin{equation}
\label{eq:mlog}
m(\varphi)=e^{i\arg{m(\pi)}}\prod_{j=1}^ne^{i\arg(\varphi-\varphi_j)}\, ,
\end{equation}
where \(2\lceil {n}/{2}\rceil\)\footnote{Here and below, $\lceil x\rceil$ stands for the 
closest integer larger than $x$ and $\lfloor x\rfloor$ for the closest integer smaller 
than $x$.}
 is the number of the discontinuities 
of $m(\varphi)$ and \(\varphi_j\) are the discontinuity points (the term  
\(2\lceil {n}/{2}\rceil\) takes into account an eventual discontinuity in $\pi$ 
that is not counted by considering the open interval $\varphi\in]-\pi,\pi[$).
We prove analytically in the next subsection that $S_\ell\propto \log \ell$ 
in the XX chain ($\gamma=0$) and then 
we show that this is not a peculiar feature of the isotropic model.

\subsection{XX chain}

\begin{figure}[t]
\begin{center}
\includegraphics[width=0.75\textwidth]{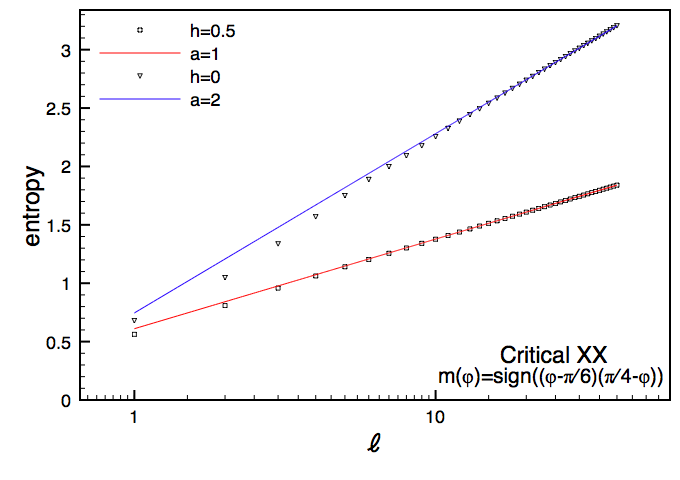}
\end{center}\caption{The entanglement entropy as a function of the block 
length for the excited state with characteristic function
$m(\varphi)={\rm sign}((\varphi-\frac{\pi}{6})(\frac{\pi}{4}-\varphi))$ 
of two critical XX chains. 
The different behavior is caused by the position of the zero modes 
($\varphi_F=\pi/6$ and $\pi/2$ with two and four discontinuities respectively)
and results in $a=1$ or $2$. 
The straight lines are the analytic prediction for large $\ell$ given 
by Eqs. (\ref{eq:criticalXX}) and (\ref{eq:addictiveconstant}).}
\label{fig:XXlog}
\end{figure}

In the XX spin chain the Bogolioubov angle reduces to  
$e^{i\theta_k}=\mathrm{sign}(J\cos\varphi_k-h)$ and the Fisher-Hartwig conjecture 
is sufficient to prove the following result: 
the entanglement entropy of the excited states described by the multi-step 
function (\ref{eq:mlog}) grows logarithmically with the width of the block. 
The coefficient in front of the logarithm is $1/6$ times the number 
of discontinuities in the non-critical region  ($|h|>1$) and it must be 
corrected in the critical region ($|h|<1$) to take into account the modes 
with zero energy. 
For $|h|<1$ the modes with zero energy at $\pm\varphi_F$ ($\varphi_F=\arccos |h/J|$)
define the function 
\begin{equation}\label{eq:mtilde}
\tilde m(\varphi)=
\cases{
m(\varphi),  &$\varphi\in[-\varphi_F,\varphi_F]$\,,\cr
-m(-\varphi),& otherwise,
}
\end{equation}
that substitutes $m(\varphi)$ when counting discontinuities.
The importance of the number of discontinuities was firstly stressed in \cite{km-04} 
in a different context.
In Fig. \ref{fig:XXlog} a direct computation shows the importance 
of the position of the modes with zero energy.

\subsubsection{Fisher-Hartwig proof of the log-behavior in the XX chain.} 
\label{App:XXproof}
The proof of the relation between the entanglement entropy and the 
discontinuities of $m(\varphi)$ when Eq. (\ref{eq:mlog}) 
holds (i.e. when $m(\varphi)=\pm1$) in an XX chain is a slight modification of 
the proof given by Jin and Korepin in Ref. \cite{jk-04} for a critical XX 
ground-state.
For $\gamma=0$, the matrix (\ref{eq:Gamma}) can be written in terms of the Pauli 
matrix \(\sigma_y\) as
\be
\Gamma(\varphi)=\pm\sigma_y m(\mp\sigma_y \varphi)\, ,
\ee
with the upper (lower) sign if the momentum $\varphi$ is below (above) the 
Fermi level of the Jordan-Wigner fermions. As a consequence the block 
Toeplitz matrix (\ref{eq:CorrelationMatrix}) can be reduced to a standard 
Toeplitz matrix with symbol
\be\fl
\gamma(\varphi)=
\cases{
1,&$\bigl(e_\varphi\!\!>0\wedge m(-\varphi)=1\bigr)\vee\bigl(e_\varphi\!\!<0\wedge m(\varphi)=-1\bigr)$,\cr
-1,&otherwise\,, 
}
\ee
with $e_\varphi=J\cos\varphi-h$. 
The reduced correlations matrix $\lambda {\bf 1}-\Pi$ is generated by the symbol
\[
t(\varphi)=\lambda-\prod_{j=1}^n e^{i\arg[\varphi-\varphi_j]}\,,
\] 
where the $\varphi_j$'s are the momenta corresponding to the $n$ 
discontinuities of $\gamma(\varphi)$. 
The ground state has two symmetric discontinuities at 
$\pm\varphi_F$. 
The symbol admits the canonical Fisher-Hartwig factorization \cite{fh}
\[
t(\varphi)=(\lambda+1)^a(\lambda-1)^b\prod_{j=1}^n t_j(\varphi)\, ,
\]
with 
\bea
t_j(\varphi)=e^{-i\beta_j(\pi-\varphi+\varphi_j)},&&
\varphi_j<\varphi<\varphi_j+2\pi,\\
\beta_j(\lambda)=\frac{(-1)^{j-1}}{2\pi i}\log\frac{\lambda+1}{
\lambda-1},\qquad &&-\pi\leq\arg\Bigl[\frac{\lambda+1}{\lambda-1}\Bigr]<\pi,
\eea
and the two exponents are 
$$b=1-a=\frac{1}{2\pi}\sum_{j=1}^n(-1)^{j-1}\varphi_j\,.$$ 

Defining $k_F\equiv\sum_{j=1}^n(-1)^{j-1}\varphi_j/2$, 
the Fisher-Hartwig conjecture (that for this case with 
$|\lambda|>1$, i.e. $|{\rm Re}(\beta_j)|<{1}/{2}$, has been proved   
by Basor \cite{fh}) reads
\bea
\fl
\det\left|\lambda\mathbf{1}-\Pi\right|\sim\prod_{i<j}^n \left[
\bigl(2-2\cos(\varphi_i-\varphi_j)\bigr)^{(-1)^{j-i}\beta(\lambda)^2}\,
\right.\\ \left.
\quad \times 
G(1+\beta(\lambda))^n
G(1-\beta(\lambda))^n\Bigl\{(\lambda+1)
\Bigl(\frac{\lambda+1}{\lambda-1}\Bigr)^{-k_F/\pi}\Bigr\}^L 
\ell^{-n\beta(\lambda)^2}\right] ,\nonumber
\eea
where \(\beta^2=\beta_j^2\) and $G(x)$ is the Barnes G-function 
\[
G(1+\beta)^nG(1-\beta)^n=e^{-(1+\gamma_E) n \beta^2}
\prod_{j=1}^\infty\Bigl(1-\frac{\beta^2}{j^2}\Bigr)^{j n} e^{n \beta^2/j}\, .
\]
In order to find the entanglement entropy we have to 
evaluate $\frac{\mathrm{d}}{\mathrm{d}\lambda}\log D_\ell(\lambda)$, where 
$D_\ell(\lambda)=\det |\lambda\mathbf{1}-\Pi|$ 
(cf. Eq. (\ref{eq:Scontour})). 
The derivative can be easily computed and it 
consists (in principle) of three terms giving in Eq. (\ref{eq:Scontour})
\be\label{eq:criticalXX}
S_{\ell}=%\frac{\rm d}{{\rm d}\la}\log D_\ell(\la)=
a_0 \ell +\frac{a}3\log\ell+a_{\{\varphi_j\}}\,,
\ee
with: 
\begin{itemize}
\item[-] the \emph{linear term} $a_0 \ell$ is the same as in the ground 
state \cite{jk-04} (except from the definition of $k_F$), 
and it is known to vanish $a_0=0$ (as actually we already proved);
\item[-] the \emph{logarithmic term} $a/3 \log \ell$ is the ground state  
contribution multiplied by $a={n}/{2}$ ($a$ will be interpreted as an effective 
central charge, that is why we multiplied by $1/3$);
\item[-] the \emph{additive constant} $a_{\{\varphi_j\}}$ is 
slightly more complicated but it has essentially the same structure of the 
ground-state and it is 
\begin{equation}
\label{eq:addictiveconstant}
a_{\{\varphi_j\}}=\frac{n}{2}a_0-
\sum_{i<j}^n \frac{(-1)^{j-i}}{6}
\log\Bigl[\sin^2\bigl(\frac{\varphi_i-\varphi_j}{2}\bigr)\Bigr] \,,
\end{equation}
and $a_0$ is the additive constant for the entanglement entropy of the 
critical XX chain without magnetic field $a_0\approx 0.726\dots$ (see 
Ref. \cite{jk-04} for the analytic expression).
Notice that it depends not only on the number of discontinuities
but also on their location.
\end{itemize}
See Fig. \ref{fig:XXlog} for a comparison of this analytic asymptotic result with the direct computation for finite $\ell$.

In Ref. \cite{ij-08} Igloi and Juhasz showed that the ground-state entropy of the XY model
with $h=0$ can be related to the sum of two Ising models (i.e. 
$\gamma=1$) with fields $h$ depending on $\gamma$. When specialized to the 
XX model, the two Ising chains are both critical and one has
\be
S_{XX}(2\ell,2N)=2S_{\rm Ising}(\ell,N)\,.
\ee
The proof in Ref. \cite{ij-08} can be generalized to some excited states by 
properly rescaling all length scales. 
Thus the knowledge of the result for the XX model, automatically gives the value 
for the critical Ising chain.
However, we do not show the details of this proof here, because in the following we will 
provide the asymptotic result for any logarithmic state of the XY chain.

At this point it is natural to wonder whether these eigenstates having an
entanglement entropy growing logarithmically with $\ell$ are the ground-states
of some conformal Hamiltonians. 
In the case of the XX model, since $H_{XX}(h)$ with different magnetic fields 
commute among each other, the ground-state at given $h$ is an excited state 
of a chain at different $h$. 
Thus, for all these states it is obvious that they should display an entanglement 
entropy scaling like Eq. (\ref{SAlog}) with $c=a=1$, i.e. they have two discontinuities in 
$\tilde m (\varphi)$. As we will see, this is true in general and
in the next subsection we show that a commuting set of local operators of 
the XY chain can be used to prove that all these logarithmic excited states are 
ground-states of properly defined local conformal Hamiltonians. 
Eq. (\ref{eq:criticalXX}) can be exploited to deduce the central 
charge of this local Hamiltonian $c=a={n}/{2}$.

\begin{figure}[t]
\includegraphics[width=0.49\textwidth]{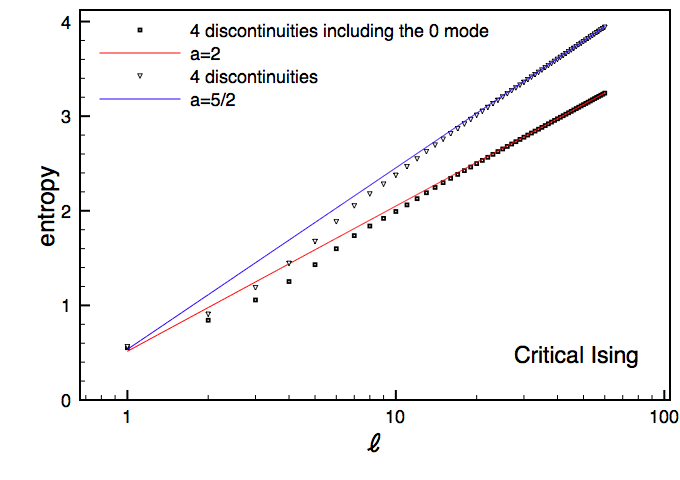}\hspace{0.01\textwidth}
\includegraphics[width=0.49\textwidth]{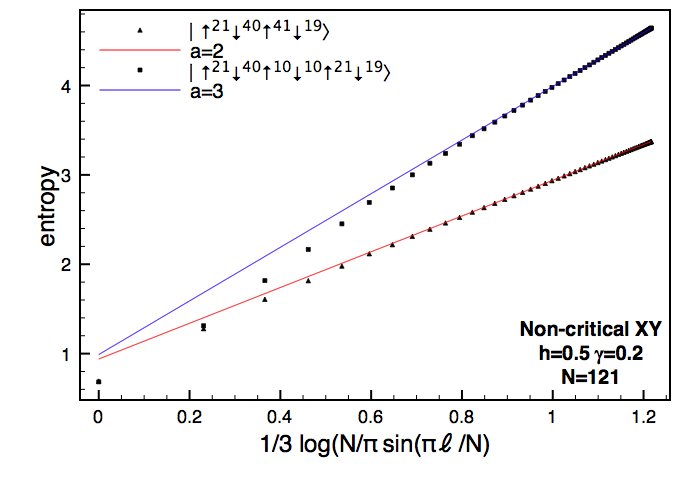}
\caption{The entanglement entropy as a function of the block length for 
two excited states of the infinite critical Ising chain with $4$ discontinuities (Left) 
at momenta $\{0,0.5,0.8,1.4\}$ and $\{-0.5,0.5,0.8,1.4\}$. 
The different slopes are caused by the zero mode.
Right: Two excited states of a non-critical XY chain in finite size.}
\label{fig:isingcancel}
\end{figure}

\subsection{Logarithmic behavior and effective Hamiltonians}
\label{ss:Non-critical systems} 

It is straightforward from Eq. (\ref{eq:Correlations}) to calculate the 
spectrum of the reduced density matrix and the entanglement entropy 
for any eigenstate at any value of $\gamma$ and $h$. 
We calculated the entanglement entropy numerically 
for several different cases and we always find a logarithmic behavior
with $\ell$ every time $m(\varphi)^2=1$ almost everywhere (see e.g. Fig. \ref{fig:isingcancel}).
To get a proof similar to the one of the previous section for the general 
XY model, one should generalize the methods in Ref. \cite{ijk-05} mapping 
the computation to a Riemann-Hilbert problem. 
This way of proceeding is very complicated and we take here a different
route based on the considerations we reported at the end of last subsection.  
In fact, this general logarithmic behavior of the entanglement entropy  
suggests that this type of excited states can be the ground states of 
critical Hamiltonians. 
We explicitly build these critical, translational invariant, and local Hamiltonians, 
proving the logarithmic behavior, with the correct prefactor.

The  excited state $|E_x\rangle$ in Eq. (\ref{EX}) is the ground state of 
all free-fermionic Hamiltonians of the form
\be\label{Heff}
\tilde H=\sum_{k}\tilde\varepsilon(\varphi_k) b^\dag_kb_k,\qquad 
{\rm with}\; \tilde\varepsilon(\varphi_k)<0 \Leftrightarrow k\in E_x,
\ee
for any choice of the function $\tilde \varepsilon(\varphi_k)$. 
In particular we could choose 
$\tilde\varepsilon(\varphi_k)=-f(\varphi_k)m(\varphi_k)$, 
with $f(x)$ an arbitrary positive function.
The choice of $\tilde \varepsilon(\varphi_k)$ determines
the locality properties of $\tilde H$: most of the choices of 
$\tilde \varepsilon(\varphi_k)$ would produce a non local $\tilde H$ (while 
by construction $\tilde H$ is always hermitian and translational invariant because it is 
built by Fourier transform).
 
To understand the locality of this effective Hamiltonian it is useful to
introduce the operators ($A^{x,y}$ are the Majorana operators introduced 
above from Ref. \cite{Vidal})
$$
G(r)=i\sum_lA_l^xA_{l+r}^y\,,\qquad  {\rm and} \qquad F^{x(y)}(r)=i\sum_lA_l^{x(y)}A_{l+r}^{x(y)}\,.
$$ 
In fact, by separating $\tilde\varepsilon(\varphi_k)$ in its even and odd part ($\tilde\varepsilon(\varphi_k)=\tilde\varepsilon_{e}(\varphi_k)+\tilde\varepsilon_{o}(\varphi_k)$),
we can rewrite the effective Hamiltonian as the sum $\tilde H=H_e+H_o$ where
\bea
\fl H_e&=&  
\sum_{r}\Bigl[\frac{1}{N}\sum_{k=\frac{1-N}{2}}^{\frac{N-1}{2}}\tilde \varepsilon_{e}(\varphi_k)e^{i\theta_k}e^{-i\varphi_k r}\Bigr]G_r\equiv
\sum_r g_e(r) G_r\,,\nonumber \\
\fl H_o&=&  i \sum_r\Bigl[\frac{1}{2N}\sum_{k=\frac{1-N}{2}}^{\frac{N-1}{2}}\tilde\varepsilon_{o}(\varphi_k)e^{-i\varphi_k r}\Bigr]\bigl(F_r^x+F_r^y\bigr)\equiv
\sum_r g_o(r) (F_r^x+F_r^y)\,,
\label{gr}
\eea
where we defined the complex couplings $g_e(r)$ and $g_o(r)$. 

The locality of $\tilde H$ is related to the long distance behavior 
of these complex couplings $g_{e/o}(r)$. 
From a standard theorem in complex analysis, we know that $g_{e/o}(r)$ decay
faster than any power (and so results in local couplings) if their Fourier transforms 
in the above equations are $C^\infty$ (i.e. with all derivatives being continuous functions;
often we will refer to these functions simply as regular).
When Eq. (\ref{eq:mlog}) holds, that is $m(\varphi)=\pm 1$ has a 
finite number of discontinuities, and for a {\it non-critical} system 
(i.e. when $e^{-i\theta_k}$ is regular), the arbitrariness in the choice of 
$\tilde \varepsilon$ allows us to take it among the $C^\infty$ functions. 
This conclude the proof for non-critical systems.

For the critical case, a slight modification is enough to give the correct Hamiltonian. 
In the XX spin chain \(e^{-i\theta}=\mathrm{sign}(J\cos\varphi-h)\) so that we can make 
the two above functions regular simply defining the characteristic function 
$\tilde m(\varphi)$
\[
\tilde m(\varphi)=\cases{
m(\varphi)  & $\varphi\in[-\varphi_F,\varphi_F]$\,,\cr
-m(-\varphi) & otherwise\, ,
}
\]
as we have already done in Eq. (\ref{eq:mtilde}).
The critical XY (\(|h|=1\)) is more involved because \(e^{-i\theta}\) can be
made regular only after imposing anti-periodic conditions to the mode of zero energy. 
It is then convenient to extend the definition of \(\tilde\varepsilon\) to 
the interval \([0,4\pi]\) 
\begin{equation}\label{eq:conditionsIsing}
\tilde\varepsilon_{(4\pi)}(\varphi)=\cases{
\tilde\varepsilon(\varphi)&$\varphi\in[0,2\pi]$\cr
-\tilde\varepsilon(4\pi-\varphi)&$\varphi\in[2\pi,4\pi]$\, .
}
\end{equation}
\(\tilde \varepsilon_{(4\pi)}\) can be chosen \(C^\infty\) because it has at 
most \(2n+2\) zeros, 
where \(n\) is the number of the discontinuities corresponding to the excited state. 
The constructed function restricted to \([0,2\pi]\) has the correct regularity properties.
Regardless of the presence of a discontinuity in \(\varphi=0\) the dispersion 
law must vanish in  \(\varphi=0\) ( see Eq. (\ref{eq:conditionsIsing})), 
thus the number of chiral modes is the number 
of discontinuities, plus \(1\) if there is not a discontinuity in \(\varphi=0\). 
This ends the construction of the local Hamiltonian for all the XY models.

And this is not yet the end of the story. We can in fact use the 
arbitrariness we have in the choice of $\tilde\varepsilon_k$ to fix it in such
a way that it crosses the zero-energy line with a non-vanishing slope. 
The low-energy properties of the resulting Hamiltonian can be then studied
by linearizing the dispersion relation close to the zeros in a canonical manner.
Each zero gives a chiral mode with central charge $1/2$ and so the total 
central charge will be $n/2$, with $n$ the number of zeros, i.e. the number 
of discontinuities of $m(\varphi)$ for non-critical systems, or the proper variation for 
critical ones (when the zero mode gives one additional contribution).  
This agrees with all the specific cases in the previous section. 
In particular if $m(\varphi)$ is discontinuous in $\varphi=0$, the zero mode contributes 
only once.
In Fig. \ref{fig:isingcancel} we report some specific examples stressing the 
importance of the critical modes and of the location of discontinuities.

\subsection{Finite size scaling}
 
When the width of the block $\ell$ is comparable with the length of the 
chain $N$, the characterization of the entanglement becomes tricky. 
When an excited state $|E_x\rangle$ can be associated to the ground-state 
of a local Hamiltonian $\tilde H$ with central charge $a=n/2$, 
i.e. when the entropy grows logarithmically with $\ell$ with a prefactor 
given by $a$, the constructive proof of previous subsection 
in the thermodynamic limit is still valid. 
Thus, in this case, the entanglement entropy has the finite size scaling 
given by Eq. (\ref{SAlog}) with $c$ replaced by $a$.
This is shown in the right panel of Fig. \ref{fig:isingcancel}.

A more intriguing problem is to understand the finite size scaling of 
excited states that have an extensive entanglement entropy in the 
thermodynamic limit. 
The result for $N\rightarrow \infty$ only predicts the derivative of the entropy for 
small subsystems. Increasing $\ell$ peculiar finite size behaviors must emerge,
because the chain is finite and the entropy must be symmetric 
around $\ell={N}/{2}$.

\begin{figure}[tbp]
\includegraphics[width=0.38\textwidth]{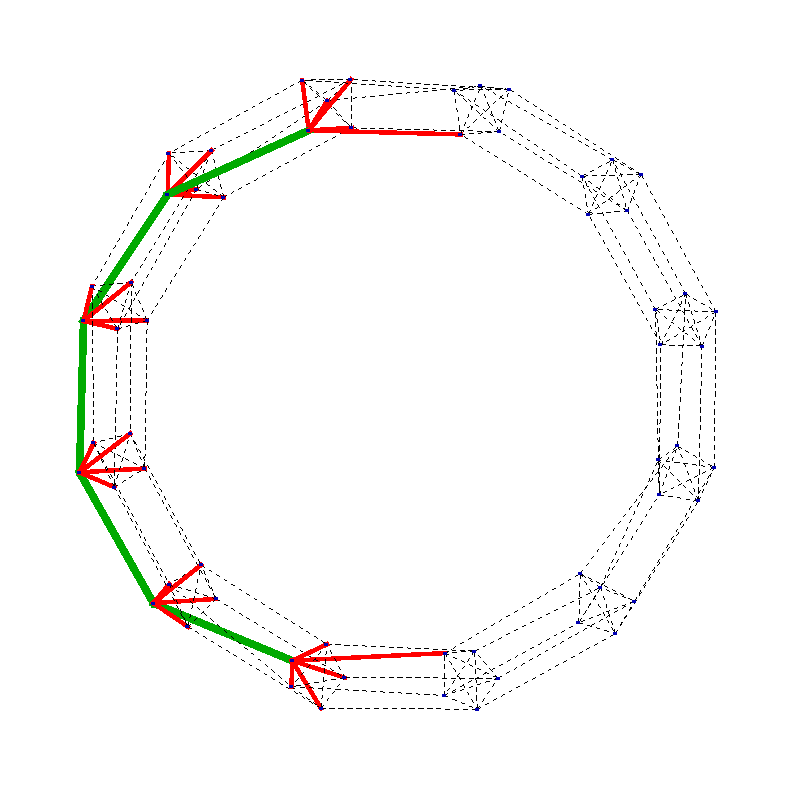}\hspace{0.96cm}
\includegraphics[width=0.38\textwidth]{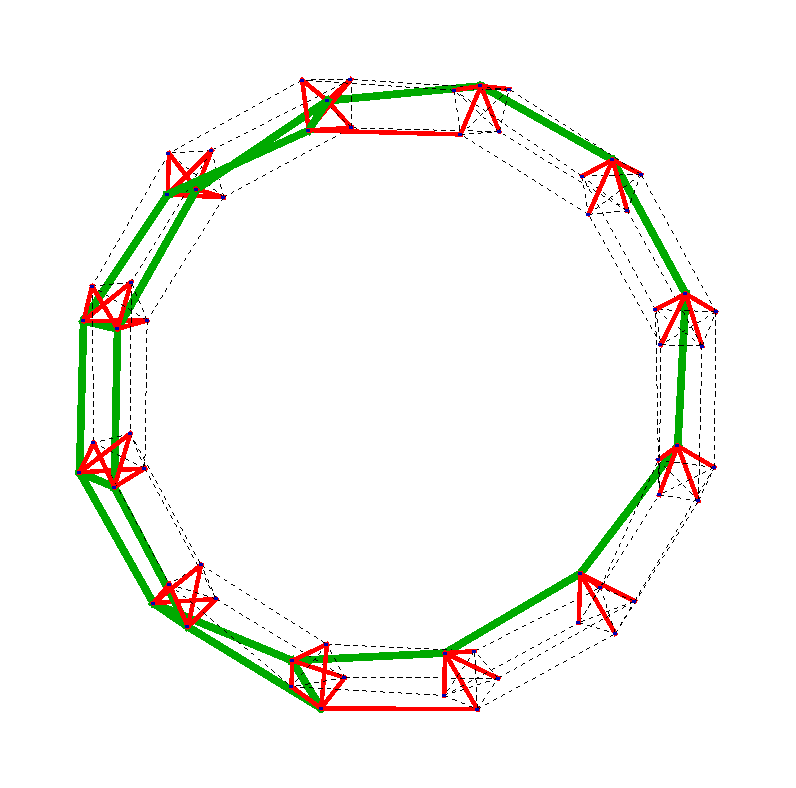}
\caption{Two 5-folded wrapped chains of 60 spins. The thick green line represents the subsystem (6 spins on the left and 18 spins on the right) while the red links give weight to the interaction between the subsystem and the rest of the chain. If the ``area law" holds the entanglement entropy is proportional to the number of the links.}
\label{fig:5fold}
\end{figure}
 
Up to now we studied in detail excited states with a regularized 
characteristic function of the type (\ref{eq:mlog}), that is 
$|{E_x}\rangle=|\prod_{j=1}^{d}\uparrow^{n_j}\downarrow^{m_j}\rangle$, 
where $n_j$ and $m_j$ are all $O(N)$ and $d$ is a finite number. 
States with $m(\varphi)^2\neq 1$ (that have extensive entanglement entropy) 
do not fall in this category as evident in the definition (\ref{mdef}). 
They can be realized by joining in a regular fashion small blocks $\kappa$ 
made by a given sequence of populated or empty energy levels 
(e.g. $\kappa=\{\uparrow\downarrow\}$ or 
$\kappa=\{\uparrow^2\downarrow\}$ etc.). 
Thus to study the finite size scaling of ``extensive'' states, we concentrate 
on those of the form 
\be
|{E_x}\rangle=| \prod_{j=1}^d \kappa ^{n_j}\bar\kappa^{m_j}\rangle\, ,
\ee
where $\bar \kappa$ is the set obtained interchanging $\uparrow$ 
with $\downarrow$. 
The entanglement entropy of this type of states in the thermodynamic limit 
has an extensive behavior because $\kappa$ averages to give 
$m(\varphi)= (u-d)/(u+d)$, 
where $u$ ($d$) is the number of up (down) arrows in $\kappa$, 
while $\bar \kappa$ gives $m(\varphi)=(d-u)/(u+d)$: 
the regularized characteristic function is a multi-step function but with 
modulus different from $1$ and Eq. (\ref{eq:extensive}) gives the 
leading term of the entanglement entropy. 

\begin{figure}[t]
\includegraphics[width=0.49\textwidth]{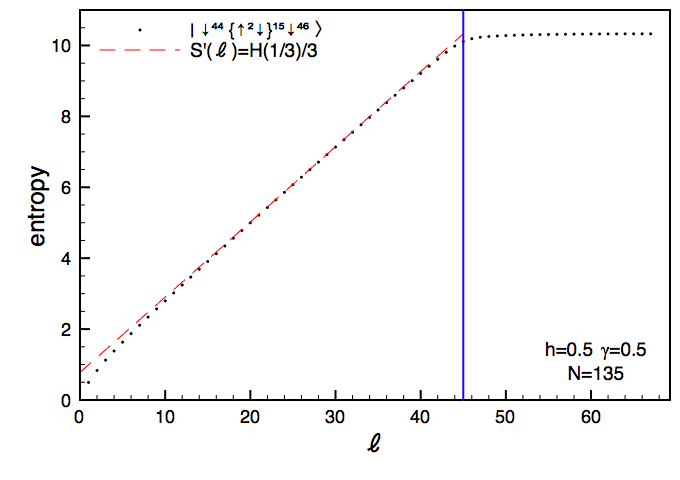}\hspace{0.012\textwidth}
\includegraphics[width=0.49\textwidth]{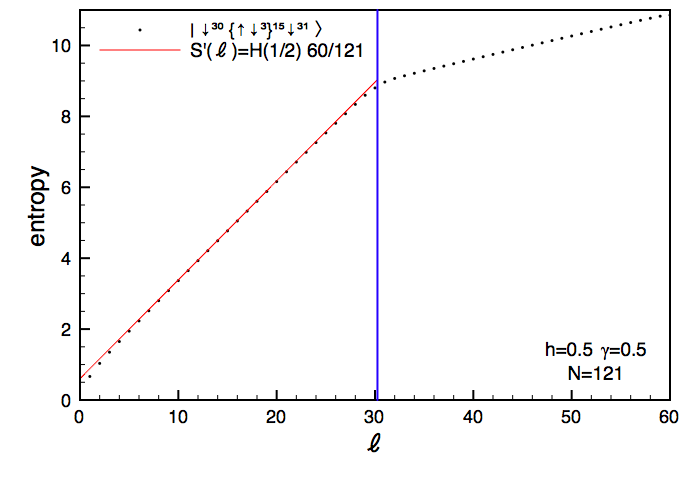}
\caption{Examples of 3- and 4-folded states.
Left: The 3-folded excited state 
$|\downarrow^{44}\{\uparrow\downarrow^2\}^{15}\downarrow^{46}\rangle$ 
for the non critical chain \((h=0.5,\gamma=0.5)\). 
Entropy grows linearly up to \({N}/{3}\) and then saturates. 
The dashed line has the slope  given by Eq. (\ref{eq:extensive}) with the 
regularized step-function \(m(\varphi)\).
Right: The 4-folded excited state 
$|\downarrow^{30}\{\uparrow\downarrow^3\}^{15}\downarrow^{31}\rangle$.
For $\ell<N/2$, the entropy always grows linearly, but with a change of slope 
close to $\ell\sim N/4$.
}
\label{fig:3fold}
\end{figure}

In order to have a quantitative prediction for the finite size scaling, 
we follow the ideas in the previous subsection by looking at the effective 
Hamiltonian obtained by the construction in Eq. (\ref{Heff}).
The resulting couplings in Eq. (\ref{gr}) could never give a finite-range Hamiltonian
because the entanglement entropy is not logarithmic.
We can make a local choice of the sign that makes  \(\tilde \varepsilon\) a regular
function (that we call $\bar \varepsilon$) giving the coupling\footnote{This coupling is a
slight modification of the ones in Eq. (\ref{gr}). It has the advantage to make all the
formulas simpler, but it applies only to non-critical systems. The generalization to 
critical ones is straightforward, but long and we do not report it here for clarity. 
However all results (except for the ground state) are independent of this choice, 
as in the previous section.}
\bea
\label{eq:g(r)}\fl
g(r)&\equiv&
\frac1N \sum_{k=\frac{1-N}2}^{\frac{N-1}2}e^{ir \varphi_k}\tilde \varepsilon(\varphi_k)\\ \fl
&=&-\frac1N\sum_{\varphi_q\in ]-\frac{\pi}{|\kappa|},\frac{\pi}{|\kappa|}[}
e^{-i |\kappa| r \varphi_q}\left[ \bar{\varepsilon}(|\kappa|\varphi_q)
+O\bigl(\frac{1}{N}\bigr)\right] 
\sum_{n=1}^{|\kappa|}\kappa_{n}\ e^{-i (n-n_0) \varphi_r}\,,\nonumber
\eea
and  the interaction is not local anymore. 
The $O(1/N)$ term comes from the series expansion of \(\bar\varepsilon\).
The first factor in equation (\ref{eq:g(r)}) is periodic of period 
${N}/{|\kappa|}$ while the second one is a modulation. 
The coupling decays faster than any power for $r<{N}/{2|\kappa|}$, but it 
explodes (i.e. it grows faster than a power) up to ${N}/{|\kappa|}$ 
when $g(r)$ becomes again of order $1$. 
The behavior for large distances is determined only by the first region
\[
g\Bigl(r+j\frac{N}{|\kappa|}\Bigr)\approx 
\frac{\sum_{n=1}^{|\kappa|} \kappa_n\ e^{-i \frac{2\pi n j}{|\kappa|}}e^{-i n \varphi_r}}{\sum_{n=1}^{|\kappa|} \kappa_n\ e^{-i n \varphi_r}}g(r)\,,
\qquad 0<r<\frac{N}{|\kappa|}\, .
\]
The interaction is localized within a distance $j N/|\kappa|$, so  
the Hamiltonian can be interpreted as a local one in a $|\kappa|$-folded 
wrapped 1D chain.    
If we assume the ``area law" to be valid for the wrapped chain (i.e. that 
only a shell of mutually interacting spins contributes to the entanglement
\cite{w-al}),
 we can predict the behavior of the entanglement entropy: 
each spin strongly interacts with the neighborhood spins and with the 
$|\kappa|$ spins  of the other wrappings (see Figure \ref{fig:5fold}). 
Thus the entanglement entropy is a piece-wise function of $\ell$ that 
changes slope at $j N/|\kappa|$. 
We can find excited states with analogous properties considering any finite \emph{partition of unity} of the circle \(]-\pi,\pi]\), with the property that all functions of the set are regular and approach step functions in the limit of large \(N\). We associate a small block \(\kappa^{(i)}\) to each function of the set and we write the coupling as a sum of terms of the form (\ref{eq:g(r)}) 
\[
g(r)=\sum_{i=1}^ng_{\kappa^{(i)}}^i(r)
\]
that we obtain identifying the regularized \(\bar \varepsilon\) with the given function of the partition. In the scaling limit the characteristic function is \(m\sim \prod_{i=1}^d \kappa_{(i)}^{n_i}\). Each \(g_{\kappa^{i}}^i\) has the behavior previously described, so the entanglement entropy is a piece-wise function of \(\ell\) changing slope in \(jN/|\kappa|\), where \(|\kappa|\) is the least common multiple of the \(\{|\kappa|^{(i)}\}\).
Two examples of 3-folded and 4-folded states are reported in Fig. \ref{fig:3fold}.

To give the details of a specific example, we report the 3-folded case $\kappa^{(1)}=\{\uparrow^2\downarrow\}$ and \(\kappa^{(0)}=\{\downarrow\}\) 
with $\varphi\in I_1\Leftrightarrow \cos\varphi\geq 1/2$ and 
$\varphi\in I_0\Leftrightarrow \cos\varphi<1/2$, 
in other words the set \(E_x\) is made of the quasiparticles with 
momenta $(2\pi(3k+q))/N$ with $|k|\leq N/12$ and \(q\in\{0,1\}\) 
$$
|E_x\rangle=\prod_{k\approx -\frac{N}{12}}^{\frac{N}{12}}b^\dag_{3k}b^\dag_{3k+1}|0\rangle\, .
$$
The excited state $|E_x\rangle$ is the ground state of the Hamiltonian 
\[
\tilde H=\sum_{k=\frac{1-N}{2}}^{\frac{N-1}{2}}\Bigl[\Bigl(\frac{1}{2}-\cos\varphi_k\Bigr)(-1)^{\lceil\frac{4k}{3}\rceil}+(-1)^{\lfloor\frac{4k}{3}\rfloor}+(-1)^{\lfloor\frac{4(k+1)}{3}\rfloor}\Bigr]b^\dag_kb_k\, ,
\]
and if \(N\) is divisible by 3 the coupling is different from \(0\) only in \(9\) points
$$
g(r)=\cases{
\frac{5}{6} & $r=0$\,,\cr
-\frac{1}{6} & $r=\pm 1$\,,\cr
-\frac{1}{6}\pm\frac{1}{\sqrt{12}}i & $r=\pm\frac{N+q}{3}\quad q\in\{-1,0,1\}$\, .
}
$$
The effective Hamiltonian \(\tilde H\) is local on the 3-folded wrapped chain.
The entropy grows linearly with the width of the block up to $\ell={N}/{3}$, after that the interaction surface does not further increase and the entanglement entropy does not 
depend anymore on the width of the block, see Figure \ref{fig:3fold} (left).
Notice on the same figure (right), the change of slope in the 4-folded case.

\begin{figure}[t]
\begin{center}
\includegraphics[width=0.49\textwidth]{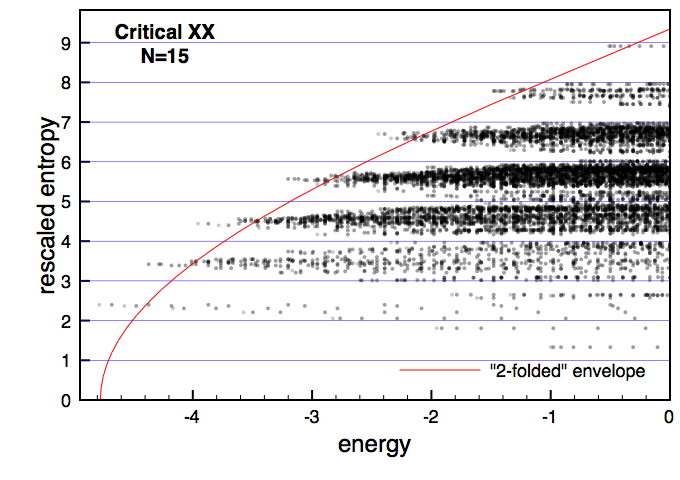}
\includegraphics[width=0.49\textwidth]{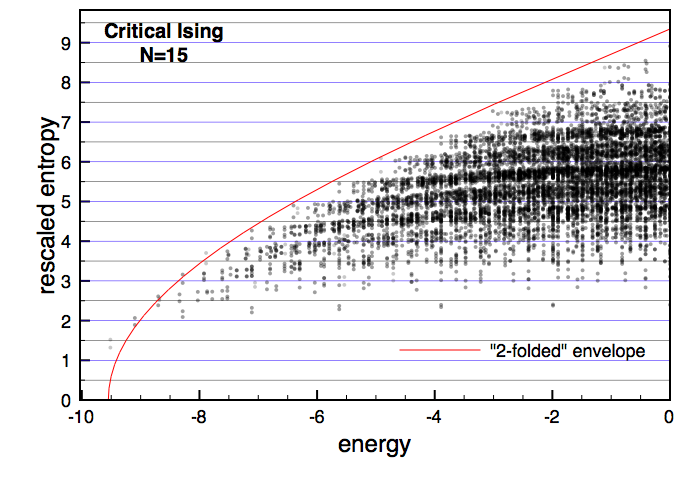}
\includegraphics[width=0.49\textwidth]{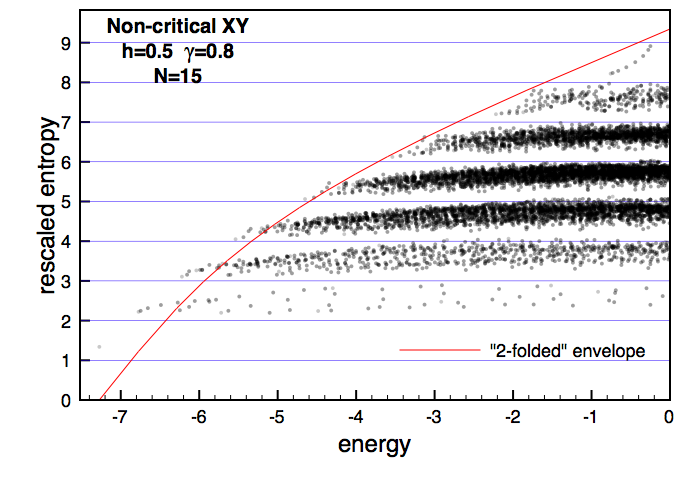}
\includegraphics[width=0.49\textwidth]{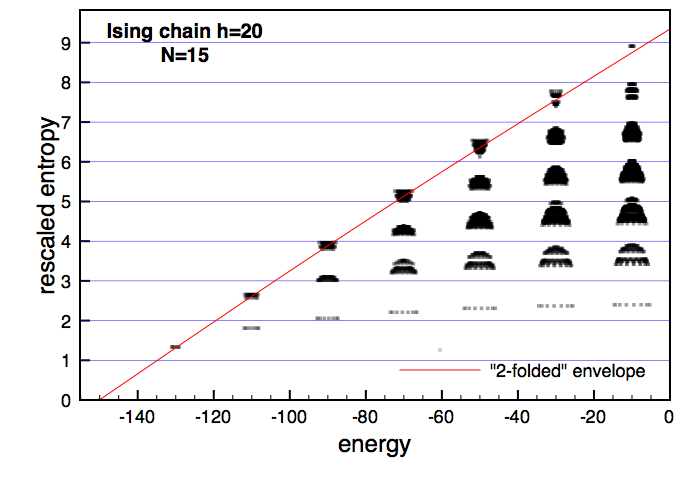}
\end{center}\caption{
Rescaled half-chain entanglement entropy $\ell=(N-1)/2$
for a critical XX in zero magnetic field, a critical Ising, a 
non critical XY spin chain, and an Ising in a very large magnetic field. 
All plots are for $N=15$. Each point corresponds to an excited state with 
energy (in unit of $J$) on the real axis.
The red curves are the ``2-folded'' estimations of the envelope.
}
\label{fig:trees}
\end{figure}

\subsection{Some general properties}

\begin{figure}[t]
\begin{center}
\includegraphics[width=0.65\textwidth]{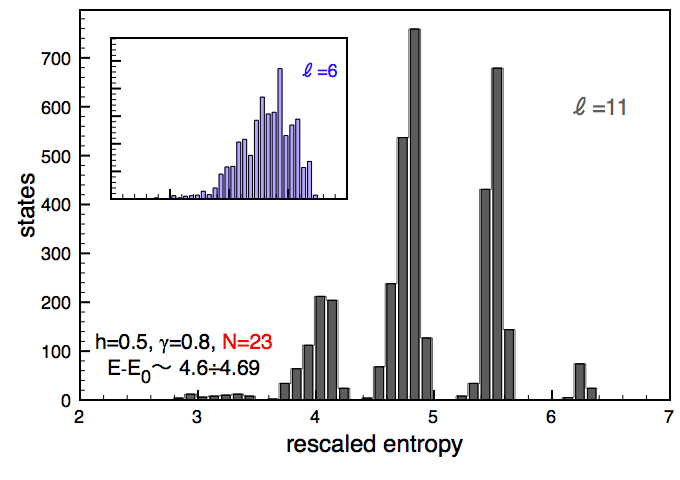}
\end{center}\caption{
Histograms for the number of the states with a given entanglement entropy
%The rescaled maximal entanglement entropy 
for a non-critical XY chain of 23 spins, after cutting the Hilbert space in an 
energy shell. 
Main plot: rescaled $S_{11}$. Inset : rescaled $S_6$.  
%In the inset is considered the entanglement entropy of a block of \(6\) spins  normalized with \(\frac{1}{3}\log\frac{23}{\pi}\sin\frac{6\pi}{23}\). 
The band-structure is evident only for $\ell=11$.}
\label{fig:freqceff}
\end{figure}

To have a general picture of the scaling of the entanglement for all excited 
states and not only in the particular classes considered so far, 
we study here the entanglement entropy in a small enough chain to be able 
to calculate it for all the $2^N$ states. We mainly concentrate on blocks 
with maximal entropy, i.e. with length equal to half-chain (actually $(N-1)/2$ 
spin, because we use $N$ odd).
Drawing general conclusions in an analytic manner for finite systems is not 
easy, so we mainly analyze numerical results. 
The plots in Fig. \ref{fig:trees} suggest that some regularities are 
general features of excited states and not only of the classes we can compute
analytically. 
In these plots (and in all those relative to this section) we always 
consider the rescaled entropy
\be\label{resc}
{\rm rescaled \; entropy}= 
\frac{S_\ell}{S_\ell^{GS}}\,,\qquad {\rm with}\;\; 
S_\ell^{GS}=\frac{1}{3}\log\bigl(\frac{N}{\pi}\sin\frac{\pi \ell}{N}\bigr)\,,
\ee 
so that, for states with a critical-like behavior (for large enough $\ell$ 
and $N$) we have a direct estimation of the effective central charge. 
We found particular instructive to plot the (rescaled) entanglement entropy 
as function of the energy of the eigenstates. 
In Fig. \ref{fig:trees}, we considered chains of 15 spins and we plot the 
rescaled $S_7$ for {\it all} the $2^{15}$ eigenstates.  
Similar plots can be done as function of total momentum instead of the 
energy.

A first feature that is particularly evident from the plots is the band-like
structure of the entanglement entropy (notice that this is independent of the 
use of the energy on the horizontal axis, any other conserved quantity would 
result in qualitative similar plots). This means that the entanglement
entropy of excited states distributes at roughly integer (or half-integer for
critical XY at $h=1$) multiples of $S_\ell^{GS}$. 
For states with a small number of discontinuities (compared to $N$), this phenomenon 
is clearly due to the quantization of the prefactor of the logarithm.
However, in general this band structure cannot be so easily explained: 
the excited states with a logarithmic behavior are expected to be negligible in 
number compared to all the others. 
Increasing the number of discontinuities at fixed $N$, the crossover to extensive 
behavior takes place and eventually it deteriorates the bands.
This last phenomenon is not evident in Fig. \ref{fig:trees} because the band structure
persists up to the maximum allowed number of discontinuities. 
The simplest explanation is that also extensive states should roughly be quantized but 
within a scale different from $S_\ell^{GS}$, that in particular does not grow with $N$.
To check this, we should increase $N$, but in doing so, the dimension of the Hilbert 
space grows exponentially and it becomes soon prohibitive to plot (and understand) 
so many points in an readable graph.
For this reason we considered a non-critical chain of $23$ spins, and to reduce
the number of states, we limited to states with energy in the interval 
$4.600<E-E_0<4.694$.  
In Fig. \ref{fig:freqceff} we report the distribution of the points. 
For $\ell=11$, the band structure is evident and the points 
distribute in an almost Gaussian fashion around some discrete values of the 
entanglement entropy, but the distance between them becomes smaller 
$S_\ell^{GS}$, confirming that the origin of this phenomenon in the upper part of the band
has nothing to do with logarithmic states.
For $\ell=6$ (inset of Fig. \ref{fig:freqceff}) the band structure disappears completely,
confirming that most of the states are extensive.
We checked that still increasing $N$, this scenario is consistent.

Another very interesting feature is that in all the plots,
the entanglement entropy has a maximum value that seems to be a regular function
of the energy (that is the final reason why we made this kind of plots). 
We argue here that these envelopes have a characteristic dependence on the 
energy that in the scaling limit is determined by excited states with extensive behavior.
We already derived the entanglement entropy for the excited states that are 
equivalent to the ground state of $n$-folded wrapped Hamiltonians. 
Eq. (\ref{eq:extensive}) characterizes the scaling regime, e.g. 
for the $2$-folded case the entanglement entropy increases linearly up to $N/2$, 
while in the $3$-folded it increases up to $N/3$ and then saturates. 
We have then for blocks of length $\ell/N \geq H(1/3)/(2H(0))=0.459\dots $ 
that the 2-folded case is more entangled than the 3-folded one.
This suggests that the 2-folded states can explain the envelopes in Fig.
(\ref{fig:trees}) for $\ell=(N-1)/2$ (Notice that the maximal entangled state, 
regardless of the energy, is always a 2-folded one).
If this is true, the envelope is easily obtained: 
the problem is analogous to find the dependence of the 
particles density on the Fermi energy in a free Fermi system at zero 
temperature. 
Indeed using Eq. (\ref{eq:extensive}) and the asymptotic expression for 
the energy, the ``2-folded approximation'' of the envelope satisfies the 
parametric equations (valid for $E<0$)
\[
\cases{
\frac{S^{MAX}}{N}\sim\frac{\log 2}{4\pi}
\int_{-\pi}^\pi\mathrm{d}\varphi\ \theta(\mu-\varepsilon)\,,\cr
\frac{E}{N}\sim-\frac{1}{4\pi}\int_{-\pi}^\pi\mathrm{d}\varphi\ \varepsilon\ \theta(\varepsilon-\mu)\, .
}
\]
In Fig. \ref{fig:trees} this analytical result is compared with the numerical 
data for a critical XX, a critical Ising and two non critical XY spin chains: 
the approximated envelope is in good agreement with the numerical data also 
for small chains. 
We also notice that 
\(\frac{\mathrm{d}S^{(MAX)}}{\mathrm{d}E}=\frac{\log 2}{\mu}\) and so  
\[
\frac{\mathrm{d}S^{MAX}}{\mathrm{d}E}\leq \frac{\log 2}{\Delta}=
\Bigl(\frac{\mathrm{d}S^{(MAX)}}{\mathrm{d}E}\Bigr)_{G.S.}\, ,
\]
where $\Delta$ is the gap in the dispersion law: if the system is critical 
then the ``2-folded'' approximation of the envelope has infinite derivative 
in \(E=E_{G.S.}\) (cf. Fig. \ref{fig:trees}).

\begin{figure}[t]
\includegraphics[width=0.49\textwidth]{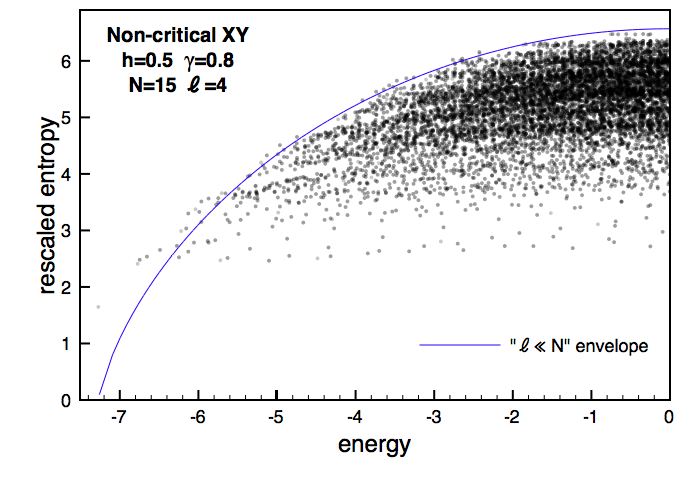}
\includegraphics[width=0.49\textwidth]{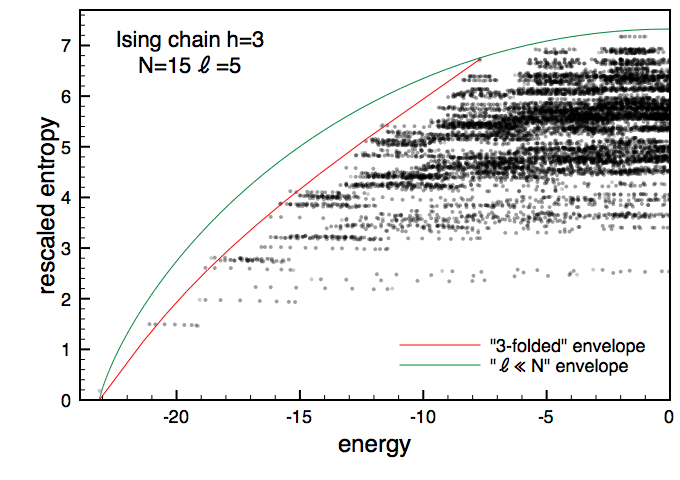}
\caption{Rescaled entanglement entropy for small blocks. 
Left:  $\ell=4$  in a non-critical XY-chain of $15$ spins; 
The continuous curve is Eq. (\ref{eq:thermalenvelop}) giving a good  
estimation of the envelope.
Right: $\ell=5$ in a non-critical Ising chain of $15$ spins; 
The ``3-folded'' envelop (in red) of the envelope is in good agreement 
with the data. For high energies, when the ``3-folded'' approximation is not 
defined, Eq. (\ref{eq:thermalenvelop}) (in green) works well.}
\label{fig:env}
\end{figure}

In the opposite limit of small $\ell$, the band structure is practically lost
(see left panel of Fig. \ref{fig:env}) and for most of the states Eq. (\ref{eq:extensive})
gives a good estimate of $S_\ell$ so that we expect that the envelope can be 
determined maximizing the expression (\ref{eq:extensive}) at fixed energy.  
The maximization gives the thermal-like parametric equations
\begin{equation}\label{eq:thermalenvelop}
\cases{
\frac{S^{MAX}}{\ell}\sim\frac{1}{2\pi}
\int_{-\pi}^\pi \mathrm{d}\varphi\ H(\tanh(\beta\varepsilon))\,,\\
\frac{E}{N}\sim -\frac{1}{4\pi}
\int_{-\pi}^\pi \mathrm{d}\varphi\  \varepsilon \tanh(\beta \varepsilon)\,,
}
\end{equation}
and the loss of the band structure can be seen as a consequence of a ``pure'' 
extensive behavior of the entropy. Eq. (\ref{eq:thermalenvelop}), in the 
scaling limit, is always an upper bound for the entanglement entropy 
because entropy is a concave function of $\ell$. 
In Fig. \ref{fig:env} (left) we compare this analytical curve
with the data for $N=15$ and $\ell=4$ in a non-critical XY-chain.

Considering blocks of intermediate lengths the parametric 
equations (\ref{eq:thermalenvelop}) define a too high bound (see right of
Fig. \ref{fig:env}). 
At the same time the band structure starts emerging.
We can improve our estimation considering a generalization of the 
``2-folded approximation'' of the envelope: the ``$n$-folded approximation'' 
(that makes sense only for $\ell\leq N/n$). 
The maximal entanglement entropy in the $n$-folded family of excited states is
\be
\cases{
\frac{S^{MAX}}{N}\sim\frac{H\bigl(1-\frac{2}{n}\bigr)}{2n\pi}\int_{-\pi}^\pi\mathrm{d}\varphi\ \theta(\mu-\varepsilon)\,,\\
\frac{E}{N}\sim\frac{1}{2n\pi}\int_{-\pi}^\pi\mathrm{d}\varphi\ \varepsilon\bigl(\theta(\mu-\varepsilon)-\frac{n}{2}\bigr)\, .
}
\ee
In Fig. \ref{fig:env} (right) we report $S_5$ for a non-critical Ising chain 
of $15$ spins (so the maximum allowed $n$ is 3). It is evident that 
up to the point where it exists the $3$-folded curve is a good 
approximation of the actual envelope, while for larger values 
Eq. (\ref{eq:thermalenvelop}) works well.

All the plots in this subsection are relative to the Slater-determinant basis. We have checked that 
considering linear combinations of eigenstates with the same energies,  these envelopes 
remain unchanged, while the band-structure disappears (as maybe could have been expected).

Lack of space prevents us to show many other similar plots about
the distribution in the energy of excited states for the entanglement entropy. 
The main features about appearance and disappearance of the band-structure and the envelopes (that we showed here with few examples) are always true.

\section{The XXZ model and the algebraic Bethe Ansatz approach to 
reduced density matrices}
\label{sec3}

We consider the anisotropic spin-$1/2$ XXZ model in the presence of a magnetic 
field in the  $z$ direction, with Hamiltonian
\begin{equation}
\label{hamh}\fl
H_\mathrm{XXZ}=\sum_{m=1}^N \Big\{ \sigma^x_m \sigma^x_{m+1} +
  \sigma^y_m\sigma^y_{m+1} + \Delta(\sigma^z_m\sigma^z_{m+1}-1) - {h \over 2} 
\sigma^z_m\Big\},
\end{equation}
and periodic boundary conditions. The model is solvable by means of the 
Bethe Ansatz for any real value of the anisotropy parameter $\Delta$
\cite{GaudinBOOK,KorepinBOOK}, 
but we will consider here only the antiferromagnetic 
critical regime $0< \Delta\leq1$ (the case $\Delta=0$ is the XX model of 
the previous section). 
We will use the quantum inverse scattering solution for this problem 
found by Kitanine, Maillet and Terras \cite{kmt-99,kmt-00}.
Some further advances for the algebraic Bethe Ansatz approach to the XXZ model
can be found in \cite{kmst-02}. 
It is worth mentioning that recently also the full solution for 
open boundary conditions has been found \cite{kkm-08}. 

The approach we follow in this paper is inspired to the ABACUS method pioneered 
by J.-S. Caux and collaborators to calculate exact dynamical correlation
functions in finite systems \cite{cm-05,cm-2,cc-06b,fcc-08,fcc-09,js-09}. 
In fact, instead of searching for exact relations valid in the 
thermodynamic limit (as e.g. made in few specific cases for the ground-state
properties \cite{ss-07,js-08,ger,sst-06}), we will work with finite chains,
solve numerically the Bethe equations for a given eigenstate and plug the
solutions in the determinant form found for the elements of the reduced 
density matrix. 
The computation of the final result will require many computational 
resources and we will discuss in the specific cases when this approach is
more convenient than exact diagonalization. 
We mention that for $\Delta=1/2$ and $N$ odd, thanks to very peculiar 
combinatorial properties \cite{rs-01}, some exact results are known also
for finite chains \cite{ncc-08,bcv-09}.

The content of next subsections is highly technical. We first review the main
results of Ref. \cite{kmt-99,kmt-00} (to make this paper self-consistent and 
to fix the notations) and then we explain the technical tricks to adapt 
these fundamental results to the computation of the reduced density matrix.
We remand the reader interested only in the results to the final subsection
\ref{Secres}.

\subsection{The algebraic Bethe Ansatz}

In the algebraic Bethe Ansatz approach (see the book \cite{KorepinBOOK} for
an introduction to the subject), the dynamics of the model is encoded in the 
so called $R$ matrix
\begin{equation}
   R(\la,\mu)=
  \left(\begin{array}{cccc}1&0&0&0\\
                 0&b(\la,\mu)&c(\la,\mu)&0\\
                 0&c(\la,\mu)&b(\la,\mu)&0\\
                 0&0&0&1 \end{array}\right),
\end{equation}
where
\[
b(\la,\mu)=\frac{\sinh(\la-\mu)}{\sinh(\la-\mu+\eta)},
\qquad c(\la,\mu)=\frac{\sinh\eta}{\sinh(\la-\mu+\eta)}.
\] 
Here the parameter $\eta$ is related to $\Delta$  by the relation
\begin{equation}
  \Delta=\frac{1}{2}(e^{\eta}+e^{-\eta}).
\end{equation}
Now we introduce the monodromy matrix
\[\fl T(\la)=R_{0 N}(\la-\xi_N)\dots 
           R_{0 2}(\la-\xi_2)R_{0 1}(\la-\xi_1)
        =\left(\begin{array}{cc}
                    A(\la)& B(\la)\\ 
                    C(\la)& D(\la)\end{array}\right),\]
where $\xi_i$ are arbitrary parameters sitting on each site of the spin chain.
The role of the inhomogeneities $\xi_i$ will become clear in the following.
We introduce the transfer matrix as trace of the monodromy matrix 
${\cal T}(\lambda)={\rm Tr}\,T(\lambda)$ that satisfies
\begin{equation}
[\lim_{\vec{\xi} \to \vec{\alpha}}{\cal T}(\lambda,\vec{\xi}),
\lim_{\vec{\xi} \to \vec{\beta}}{\cal T}(\lambda,\vec{\xi})]=0,\qquad 
{\rm with} %\;\; \vec{\alpha}\, \textrm{of the form}
\quad \vec{\alpha}=(\alpha,\dots,\alpha)\;, %\forall \alpha\,,
\end{equation}
where we denoted with $\vec{\xi}$ the vector with components $\xi_i$.
In the approach of Ref. \cite{kmt-99} keeping the $\xi_i$ different 
helps in deriving general results. 
Only at the end, to recover the results for the XXZ model, we will take 
the so called homogeneous limit $\vec{\xi}\to\vec{\alpha}$.
Every eigenstate of the Hamiltonian (\ref{hamh}) can be written as
\be
\ket{\{\lambda_i\}} =\prod_{k=1}^M B(\la_k) \ket{0}, \qquad
\bra{\{\lambda_i\}} =\bra{0} \prod_{k=1}^M C(\la_k),\\
\label{states}
\ee
where we denoted with $\ket{0}$ the reference state with all spins up
\begin{equation}\label{refstate}
\ket{0}=\bigotimes_{k=1}^N\ket{+}_k\,.
\end{equation}
The parameter $M$ is such that $M\le N/2$ and $M=N/2$ for the ground state in zero 
magnetic field, 
while the parameters $\{\lambda_1,\dots,\lambda_M\}$ are called rapidities.
We also introduce $d(\lambda)$ 
\begin{equation}
d(\lambda)=\prod_{i=1}^N b(\lambda,\xi_i)\,, \qquad
{\rm for\; which}\;\;
d(\xi_i)=0\quad\forall i\,.
\label{supp}
\end{equation}
Not all states of the form Eq. (\ref{states}) are eigenstates of the Heisenberg Hamiltonian:
the rapidities $\lambda_i$ must satisfy a set of non-linear equation known as 
{\it Bethe equations} that for the Heisenberg chain can be written as 
\begin{equation}
   \frac{1}{d(\lambda_j)}\prod\limits_{k=1\atop k \not= j}^M
      \frac{b(\lambda_j, \lambda_k)}{b(\lambda_k, \lambda_j)} 
   = 1,                    \qquad 1 \le j \le M. 
\label{bethe}
\end{equation}

We need the commutation relations
\be
\big[ B(\lambda),B(\mu) \big] =
\big[ C(\lambda),C(\mu) \big] =0, \qquad{\rm for\;all}\; \lambda,\mu\,,
\label{commut}
\ee
to derive the action of the operators $A\,,B\,,C\,,D$ on an arbitrary state
$\ket{\{\lambda_i\}}$ \cite{kmt-00} 
\bea
\fl
 \bra{0}\pl_{k=1}^M C(\la_k)\,A(\la_{M+1})&=&
  \sul_{a'=1}^{M+1} a(\la_{a'})
   \frac{\pl_{k=1}^M
   \sinh(\la_k-\la_{a'}+\eta)}{\pl_{k=1\atop{k\neq a'}}^{M+1}\sinh(\la_k-\la_{a'})}
   \,\bra{0}\pl_{k=1\atop{k\neq a'}}^{M+1} C(\la_k);\label{abbb}\\
\fl \bra{0}\pl_{k=1}^M C(\la_k)\, D(\la_{M+1})&=&
  \sul_{a=1}^{M+1} d(\la_a)
   \frac{\pl_{k=1}^M
   \sinh(\la_a-\la_k+\eta)}{\pl_{k=1\atop{k\neq a}}^{M+1}\sinh(\la_a-\la_k)}
   \,\bra{0}\pl_{k=1\atop{k\neq a}}^{M+1} C(\la_k),
   \label{dbbb}\\
\fl \bra{0}\pl_{k=1}^M C(\la_k)\, B(\la_{M+1})&=&\sul_{a=1}^{M+1}
  d(\la_a)\frac{\pl_{k=1}^M
   \sinh(\la_a-\la_k+\eta)}{\pl_{k=1\atop{k\neq a}}^{M+1}\sinh(\la_a-\la_k)}
\times  \nonumber \\&&
\fl \sul_{a'=1\atop{a'\neq a}}^{M+1} \frac{a(\la_{a'})}{\sinh(\la_{M+1}-\la_{a'}+\eta)}
\frac{\pl_{j=1\atop{j\neq a}}^{M+1}
  \sinh(\la_j-\la_{a'}+\eta)}{\pl_{j=1\atop{j\neq a,a'}}^{M+1}\sinh(\la_j-\la_{a'})}
\bra{0} \pl_{k=1\atop{k\neq a,a'}}^{M+1}]  C(\la_k)\,.
\end{eqnarray}
We can fix $a(\lambda)=1$ for all $\lambda$. 

A fundamental ingredient is the formula for the scalar product between two arbitrary states. 
Given a set $\{\lambda_1,\dots,\lambda_M\}$ that is solution to the Bethe equations
(\ref{bethe}) and another set of arbitrary numbers $\{\mu_1,\dots,\mu_M\}$, 
the scalar product of states of the form (\ref{states}) is given by the so 
called Slavnov formula \cite{s-89} 
\begin{eqnarray}
\fl \bra{0}\ \prod_{j=1}^M C (\mu_j) \
\prod_{k=1}^M B (\lambda_\alpha) \ket{0}=
\frac {\det H(\{\la_\a\},\{\mu_j\})}
  {\pl_{j>k}
  \sinh(\mu_k-\mu_j)
     \pl_{\alpha<\beta} \sinh(\lambda_{\b}-\lambda_{\a})},
\label{slav}
\end{eqnarray}
where we defined
\begin{equation}
\fl  H_{a b}=\frac{\sinh(\eta)}{\sinh(\la_a-\mu_b)}
  \Bigg( \frac{1}{d(\mu_b)}\pl_{m\neq a}\sinh(\la_m-\mu_b+\eta)
     -\pl_{m\neq a}\sinh(\la_m-\mu_b-\eta) \Biggr).
\label{matrslav1}
\end{equation}
When $\{\lambda_i\}=\{\mu_i\}$, Eq. (\ref{slav}) gives the Gaudin formula for 
the norm of a Bethe state \cite{GaudinBOOK,KorepinCMP86}
\begin{equation}
\fl \langle 0|\prod_{j=1}^{M}C(\lambda_j)
\prod_{j=1}^{M}B(\lambda_j)|0\rangle=\sinh^M\eta
\prod\limits_{a,b=1\atop{a\ne b}}^{M}
\frac{\sinh(\lambda_a-\lambda_b+\eta)}
{\sinh(\lambda_a-\lambda_b)}{\det}_M H'(\{\lambda\}).
\end{equation}
where $H'$ is
\begin{equation}
H'_{jk}(\{\lambda\})
=-\delta_{jk}\left[\frac{d'(\lambda_j)}{d(\lambda_j)}
-\sum_{a=1}^{M} K(\lambda_j-\lambda_a)\right]-K(\lambda_j-\lambda_k),
\end{equation}
and 
\begin{equation}
K(\lambda)=\frac{\sinh(2\eta)}{\sinh(\lambda+\eta)\sinh(\la -\eta)}\,.
\end{equation}
Notice that for the following manipulations, it is fundamental that the set of numbers
$\{\mu_i\}$ could not be solution of some Bethe equations.

\subsection{Reduced density matrix}

Let us consider a given Bethe state $\ket{\{\lambda_i\}}$, and let us
select a block of length $\ell$ as a subsystem of the spin chain. 
Every element of the reduced density matrix of these $\ell$ contiguous spins
can be written as
\be
P^{\epsilon'_1,\dots ,\epsilon'_\ell}_{\epsilon_1,\dots ,\epsilon_\ell}\equiv
\frac{\bra{\Psi} E_1^{\epsilon'_1\epsilon_1} \cdots E_l^{\epsilon'_\ell\epsilon_\ell} 
\ket{\Psi}}{\langle \Psi\arrowvert \Psi\rangle}\,,
\label{dm}
\ee
where the indices $\epsilon$ can have the values $\{+,-\}$ and 
the matrices $E^{\epsilon,\epsilon'}$ are
\bea\fl
E^{++}_j=\left(\begin{array}{cc}
1&0\\0&0\\\end{array}\right)_{\!\![j]}=\frac12+S_j^z, \qquad&&
E^{--}_j=\left(\begin{array}{cc}
0&0\\0&1\\\end{array}\right)_{\!\![j]}=\frac12-S_j^z, \nonumber\\ \fl
E^{+-}_j=\left(\begin{array}{cc}0&1\\0&0\\\end{array}\right)_{\!\![j]}=
S_j^x+i S_j^y, \qquad&&
E^{-+}_j=\left(\begin{array}{cc}0&0\\1&0\\\end{array}\right)_{\!\![j]}
=S_j^x-i S_j^y. \nonumber
\eea
Once we know the reduced density matrix, any multi-point 
correlation function built within the $\ell$ spins can be found by 
considering the appropriate linear combinations. 
The most general object we need is
\begin{equation}
F_\ell(\{\e_j,\e'_j\})=\frac {\bra{\Psi}\pl_{j=1}^\ell
E^{\e'_j,\e_j}_j\ket{\Psi}}{\bra{\Psi}\Psi\rangle}\,,
\label{genabcd0}
\end{equation}
that has been obtained in Ref. \cite{kmt-00}
\begin{equation}
F_\ell(\{\e_j,\e'_j\})=\phi_\ell(\{\la\})\frac {\bra{\Psi}
T_{\e_1,\e'_1}(\xi_1)\dots T_{\e_\ell,\e'_\ell}(\xi_\ell)\ket{\Psi}}{\bra{\Psi}\Psi\rangle},
\label{genabcd}
\end{equation}
where 
\[\phi_\ell(\{\la\})=\pl_{j=1}^\ell\pl_{a=1}^M
\frac{\sinh(\la_a-\xi_j)}{\sinh(\la_a-\xi_j+\eta)}.\]
Before reporting the main result of \cite{kmt-00}, 
we have to define the following two sets of indices
\bea
\mathbf{\alpha^+}=\{j:\, 1\le j\le \ell, \,\e_j=+\}\,,\\
\mathbf{\alpha^-}=\{j:\, 1\le j\le \ell,\, \e'_j=-\}\,.
\eea
We denote with $d^+$ ($d^-$) the dimension of the set $\alpha^+$ ($\alpha^-$).
For each $j\in\mathbf{\alpha^\pm}$ it is necessary to define a set $a_j$ 
(if $j\in\mathbf{\alpha^-}$) and a set $\a'_j$ (if $j\in\mathbf{\alpha^+}$) 
such that 
\[1\le a_j\le M+j, \,\,
a_j\in \mathbf{A}_j,\quad 1\le a'_j\le M+j, \,\, a'_j\in \mathbf{A'}_j.\]
where we introduced 
\bea
\mathbf{A}_j=&\{b: 1\le b\le M+\ell, \,\,b\neq a_k,a'_k,\,\, k<j\},\\
\mathbf{A'}_j=&\{b: 1\le b\le M+\ell, \,\,b\neq a'_k, \,k<j,\,
b\neq a_k, \,k\le j\}.
\eea
Now we need only the redefinition
\begin{equation}
\{\la_k\}\rightarrow\{\la_k,\xi_1\dots\xi_\ell\}\,,
\label{ridef}
\end{equation}
to write \cite{kmt-00}
\begin{eqnarray}\fl
\bra{0}\pl_{k=1}^M C(\la_k)\,T_{\e_1,\e'_1}(\la_{M+1})\dots
T_{\e_\ell,\e'_\ell}(\la_{M+\ell})=
\label{action}
\\
\nonumber\qquad\qquad
\sul_{\{a_j,a'_j\}}G_{\{a_j,a'_j\}}(\la_1,\dots,\la_{M+\ell})\bra{0}\pl_{b\in \mathbf{A}_{l+1}}
C(\la_b)\,,
\end{eqnarray}
where 
\bea\fl
G_{\{a_j,a'_j\}}(\la_1,\dots,\la_{M+\ell})=\pl_{j\in\mathbf{\a^-}}d(\la_{a_j})&
\frac{\pl_{b=1\atop{b\in\mathbf{A}_j}}^{M+j-1}\sinh(\la_{a_j}-\la_b+\eta)}
{\pl_{b=1\atop{b\in\mathbf{A'}_j}}^{M+j}
          \sinh(\la_{a_j}-\la_b)}
\times\nonumber\\&\times
\pl_{j\in\mathbf{\a^+}}
\frac{\pl_{b=1\atop{b\in\mathbf{A'}_j}}^{M+j-1
                                                }\sinh(\la_b-\la_{a'_j}+\eta)}
{\pl_{b=1\atop{b\in\mathbf{A}_{j+1}}}^{M+j}\sinh(\la_b-\la_{a'_j})}.
\label{funabcd}
\eea
An important simplification comes from the relation (\ref{supp}) which 
allows 
$$
a_j\le M\quad \forall j\,.
$$

Now it is important to know {\it how many terms are involved in the summation} 
in Eq.  (\ref{action}).
By simple counting we get 
\be
\frac{M!}{(M-d^-)!}\prod_{i=1}^{d^+}\big(M-d^-+\alpha^+_i-i+1\big)\,,
\ee
that in the limit of large $M$ behaves as $M^{\ell}$.

We stress now one of the main features of this approach 
for the calculation of the reduced density matrix. 
The computational resources we need for the algorithm grow exponentially with 
$\ell$, so it would be comparable to exact diagonalization, but at fixed 
$\ell$ they only grows algebraically with $M$ (but with a power equal to $\ell$).
Thus we can expect that for relative small $\ell$ we can calculate the
reduced density matrix for very large systems, while exact diagonalization 
can work with at most about $30$ spins. 

Thanks to the invariance under permutations of the set $\{\la_1,\dots,\la_k\}$ 
in the Slavnov formula, the number of determinants we need to calculate 
can be reduced to
\begin{eqnarray}
\sum_{i=1}^{d^+}\bigg\{\sum_{j_1<j_2<\dots<j_i}^{d^+}\bigg[\prod_{k=1}^i(\alpha_{j_k}-\alpha_{j_k-1})\bigg]
\left(\begin{array}{c} M\\{d^-+d^+-i}\end{array}\right)\bigg\}
\end{eqnarray}
We exploited this symmetry to reduce the computational effort.

To proceed further, we define the set $\mathcal{A}$
\be
\mathcal{A}=\{(a_1,\dots,a_{d^-+d^+})\}\,,
\ee
whose elements have the property that $a_i\ne a_j\,\forall i,j$ and
\begin{eqnarray}
a_i\in\{1,\dots, M\} & \quad\textrm{if}\quad 1\le i\le d^-\,,\\
a_i\in\{1,\dots,M+\alpha^+_i\} &\quad\textrm{if}\quad d^-< i\le d^-+d^+\,,
\end{eqnarray}
that allows to  rewrite the summation in Eq. (\ref{action})
as a sum over the elements of the set $\mathcal{A}$
\begin{equation}
\sum_{\{a_j,a'_j\}}=\sum_{\{a_1,\dots,a_{d^-+d^+}\}\in \mathcal{A}}\,.
\end{equation}
Using the definition of $\mathbf{A}_j$, we can write 
\begin{equation}
\bra{0}\pl_{b\in \mathbf{A}_{l+1}} C(\la_b)=
\{\lambda_{i_1},\lambda_{i_2},\dots,\lambda_{i_{M-n}},\xi_{j_1},\dots,\xi_{j_n}\}
\label{rearr}
\end{equation}
where $1\le i_i\le M$ and $1\le j_i\le l$.
We stress that in general $n\ne \ell$.

In order to calculate (\ref{genabcd}) we have to take the scalar product 
between (\ref{rearr}) and the state
\begin{equation}
\pl_{b=1}^{M} B(\la_b)\ket{0}\,.
\label{state2}
\end{equation}
Using again (\ref{commut}) we can rearrange the set of $\lambda_i$
\begin{equation}
\{\lambda_{i_1},\lambda_{i_2},\dots,\lambda_{i_{M-n}},\lambda_{k_1},
\dots,\lambda_{k_n}\}\,,
\end{equation}
where $1\le k_i\le M$.
The important point is that the first $M-n$ rapidities  $\lambda_i$ are the same 
as in Eq. (\ref{rearr})
and give a trivial contribution to the scalar product.

\subsection{Homogeneous limit}

The tricky task in dealing with algebraic Bethe Ansatz is to 
take the homogeneous limit in Eq. (\ref{action}).
In order to perform this limit, we consider first the Slavnov formula. 
We remind that in general we have to do the scalar product between the
two states
\begin{eqnarray}
\nonumber
\ket{\{\mu_i\}}=|\{\lambda_{i_1},\lambda_{i_2},\dots,\lambda_{i_{N-n}},\xi_{j_1},\dots,\xi_{j_n}\}\rangle\,,\\
\nonumber
\ket{\{\lambda_i\}}=|\{\lambda_{i_1},\lambda_{i_2},\dots,\lambda_{i_{N-n}},\lambda_{k_1},\dots,\lambda_{k_n}\}\rangle\,,
\end{eqnarray}
where $0\le n\le \ell$ and $i_s,k_s$ can take values in the interval $[1,M]$ 
while $j_s\in[1,\ell]$.
We have to specialize Eq. (\ref{slav}) to  this case. 
It is easy to show that 
\begin{eqnarray}
\bra{\{\lambda_i\}}\{\mu_j\}\rangle =\frac{\sinh(\eta)^M\pl_{j=1}^M\pl_{i=1}^M
\sinh(\lambda_i -\mu_j +\eta )}
{\pl_{i=1}^M\pl_{j<i}\bigg[\sinh(\la_i - \la_j)\sinh(\mu_j - \mu_i)\bigg]}
\det T\,,
\label{slav1}
\end{eqnarray}
where we introduced the matrix $T$ whose elements are
\begin{equation}
T_{ij}=\left\{
\begin{array}{lr}
H'_{ij} & j\le M-n\,,\\ \displaystyle
 -\frac{1}{\sinh(\lambda_a-\mu_b)\sinh(\lambda_a-\mu_b+\eta)} & 
\qquad j> M-n\,.
\end{array}
\right.
\end{equation}
Recalling that for $j>M-n$ we have $\mu_i=\xi_{j_i}$, it follows that in the 
homogeneous limit we have a matrix whose last $n$ rows are equal, so the
determinant is zero. This is compensated by the prefactor 
$\pl_{i=1}^M\pl_{j<i}\sinh(\mu_j - \mu_i)$ in Eq. (\ref{slav1}) that is
vanishing. 

To obtain the finite result of this limiting procedure let us define
\begin{equation}
f_i(x)=f(\la_i,x)=
\frac{1}{\sinh(\lambda_i-\frac{\eta}{2}-x)\sinh(\lambda_i+\frac{\eta}{2}-x)}\,.
\end{equation}
For an arbitrary value of $n$ (ignoring for the moment the minus signs) 
we have to take the determinant of the matrix
\begin{equation}
T_{ij}=\left (
\begin{array}{cccc}
 &  \textrm{regular}&  &  \\
 &  \textrm{terms}&  &  \\
\hline
f_1(\epsilon_{j_1}) & f_2(\epsilon_{j_1}) & \dots & f_M(\epsilon_{j_1})\\
f_1(\epsilon_{j_2}) & f_2(\epsilon_{j_2}) & \dots & f_M(\epsilon_{j_2})\\
\vdots & \ddots& &\vdots\\
f_1(\epsilon_{j_n}) & f_2(\epsilon_{j_n}) & \dots & f_M(\epsilon_{j_n})
\end{array}
\right )\,,
\label{matrix0}
\end{equation}
where
\begin{equation}
\xi_{j_i}=\frac{\eta}{2}+\e_{j_i}\,.
\end{equation}
Let us now consider the Taylor expansion of the $f(a,x)$ around $x=0$
\begin{equation}
\fl f(a,x)=f(a,0)+f'(a,0)x+\frac{1}{2!}f^{''}(a,0)x^2+\dots +\frac{1}{n!}f^{(n)}(a,0)x^n +\dots\,.
\end{equation}
Gauss manipulations on the matrix above give
the same result on each column except the index $a$ which distinguishes 
the different columns.
Therefore  we can restrict to one column and construct the
following matrix
\begin{equation*}
\left(
\begin{array}{ccccc}
f_i(0) & f'_i(0)\epsilon_{j_1} & \dots & \frac{1}{\ell!}f^{(\ell)}_i(0)\epsilon_{j_1}^\ell\dots\\
f_i(0) & f'_i(0)\epsilon_{j_2} & \dots & \frac{1}{\ell!}f^{(\ell)}_i(0)\epsilon_{j_2}^\ell\dots\\
\vdots & \ddots& &\vdots\\
f_i(0) & f'_i(0)\epsilon_{j_n} & \dots & \frac{1}{\ell!}f^{(\ell)}_i(0)\epsilon_{j_n}^\ell\dots
\end{array}
\right )\,.
\end{equation*}
Since each column of the last matrix  has the term $f^{(k)}_i(0)$ we can 
neglect it (we will restore
it at the end of the manipulations) and consider the matrix
\begin{equation}
\left(
\begin{array}{cccccc}
1 & \epsilon_{j_1} & \frac{1}{2!}\epsilon_{j_1}^2  & \dots & \frac{1}{\ell!}\epsilon_{j_1}^\ell\dots\\
1 & \epsilon_{j_2} & \frac{1}{2!}\epsilon_{j_2}^2  & \dots & \frac{1}{\ell!}\epsilon_{j_2}^\ell\dots\\
\vdots & \ddots& & & \vdots\\
1 & \epsilon_{j_n} & \frac{1}{2!}\epsilon_{j_n}^2  & \dots & \frac{1}{\ell!}\epsilon_{j_n}^\ell\dots
\end{array}
\right )\,.
\label{matrix}
\end{equation}
By mean of rows manipulations it is possible to put the last matrix in 
a triangular form
\begin{equation}
\left(
\begin{array}{ccccc}
1 & g_1(\e_{j_1}) & \frac{1}{2!}g_1(\e_{j_1}^2) & \dots & \frac{1}{\ell!}g_1(\e_{j_1}^\ell)\\
0 & g_2(\e_{j_1},\e_{j_2}) & \frac{1}{2!}g_2(\e_{j_1}^2,\e_{j_2}^2) & \dots & \frac{1}{\ell!}g_2(\e_{j_1}^\ell,\e_{j_2}^\ell)\\
0 & 0                      & \frac{1}{2!}g_3(\e_{j_1}^2,\e_{j_2}^2,\e_{j_3}^2) &\dots & \frac{1}{\ell!}g_3(\e_{j_1}^\ell,\e_{j_2}^\ell,\e_{j_3}^\ell)\\
0 & 0                      & \ddots & &
\end{array}
\right )\,,
\label{matrix1}
\end{equation}
where we have for instance $g_1(x)=x$, $g_2(x,y)=x-y$ and more complicated 
expressions for the other functions. 
In order to obtain the homogeneous limit in the Slavnov formula, we need one 
more step: since we know that the limit exists, we have to  choose in a 
convenient way the variables $\e_{j_i}$. One possible choice is
\begin{equation*}
\xi_j=\left\{
\begin{array}{lr}
\frac{\eta}{2} & j=1\\
\frac{\eta}{2}+\e\exp(\frac{2\pi i}{n-1}j) & j>1
\end{array}
\right.
\end{equation*}
that corresponds to $\e_{j_k}=\e\exp(\frac{2\pi i}{n-1}j_k)$.
It is useful to consider the simpler case in which
$\e_{j_i}=\e_i$. 
Substituting in (\ref{matrix}) we obtain that the matrix (\ref{matrix1}) 
has a simple form. Indeed it is easy to see that (\ref{matrix1}) becomes 
proportional to the identity matrix (up to the $n$-th order) $K \mathbf{1}$ , with
\begin{equation}
K=\det\bigg[\frac{\exp(\frac{2\pi i}{n-1}jk)}{j!}\bigg]_{j,k}\,.
\end{equation}
In conclusion this means that to have the lowest order in $\e$ for the 
Slavnov determinant we can write the matrix (\ref{matrix0}) as
\begin{equation}
T_{ij}=\left\{
\begin{array}{lr}
H'_{ij} & j\le M-n\,,\\
f_i(0)  & j=M-n+1\,,\\
Kf^{(j-M+n-1)}_i(0) & j>M-n+1\,.
\end{array}
\right.
\label{sub}
\end{equation}
Moreover we have to consider the contribution given by 
\begin{equation}
\pl_{j,k=1\atop{j>k}}^n\sinh(\xi_k-\xi_j)= %(-1)^n
(-\epsilon)^{{n(n-1)}/{2}}\pl_{k=1\atop{j>k}}^{n-1}
\bigg(e^{i \frac{2\pi j}{n-1}}-e^{i \frac{2\pi k}{n-1}}\bigg)\,.
\end{equation}
This concludes the calculation of the homogeneous limit in the Slavnov formula.

We can now ask {\it how many terms is it possible to obtain with the algorithm 
developed so far}.
The answer can be given examining the function $G$ in Eq. (\ref{action}). 
If the function $G$ has no poles, then the procedure just outlined 
works with no modification.
Unfortunately this happens only in very few cases, for example for 
the first element of the reduced density matrix, that is the so called 
emptiness formation probability
 \begin{equation}
\tau(\ell)=\frac {\bra{\psi_g}\pl_{j=1}^\ell
\frac 12(1-\sigma^z_j)\ket{\psi_g}}{\bra{\psi_g}\psi_g\rangle}\,.
\end{equation}
(On passing it is worth mentioning that this element can be computed 
in the thermodynamic limit \cite{kefp,cp-08,KorepinBOOK}, basically because of this 
simplification.)
In this case Eq. (\ref{action}) simplifies to
\begin{equation}
\tau(\ell)=\phi_\ell(\{\la\})
\frac{\bra{0}\pl_{a=1}^M C(\la_a)\pl_{j=1}^\ell D(\xi_j)\pl_{a=1}^M B(\la_a)\ket{0}}
{\bra{0}\pl_{a=1}^M C(\la_a)\pl_{a=1}^M B(\la_a)\ket{0}},
\label{efpddd}
\end{equation}
that can be written as
\begin{eqnarray}
\bra{0}\pl_{k=1}^M C(\la_k)\pl_{j=1}^\ell
D(\la_{M+j})=\sul_{a_1=1}^{M+1}
\sul_{a_{2}=1\atop{a_{2}\neq a_1}}^{M+2}\dots\!\!\\
\nonumber
\dots\sul_{a_{l}=1\atop{a_{l}\neq a_1,\dots,a_{l-1}}}^{M+\ell}
G_{a_1\dots a_l}(\la_1\dots\la_{M+\ell})
\bra{0}\pl_{k=1\atop{k\neq a_{1},\dots,a_l}}^{M+\ell} C(\la_k),
\label{efpsum}
\end{eqnarray}
where $G$ is
\begin{equation}\fl
G_{a_1\dots a_\ell}(\la_1,\dots\,la_{M+\ell})=\pl_{j=1}^\ell
d(\la_{a_j})
\frac{\pl_{b=1\atop{b\neq a_{1},\dots,a_{j-1}}}^{M+j-1}\sinh(\la_{a_j}-\la_b+\eta)}
{\pl_{b=1\atop{b\neq a_1,\dots,a_j}}^{M+j}
          \sinh(\la_{a_j}-\la_b)}.
\label{G}
\end{equation}
By definition $G$ cannot diverge, then all the machinery developed so far is 
enough to compute $\tau(\ell)$.
In Table \ref{efp} we report some results for the emptiness formation 
probability for a chain of length $L=20$ at $\Delta=0.5$
in the ground state. 
The agreement with DMRG data is perfect (taking into account the numerical 
rounding-off).

\begin{table}[t]
\centering
\begin{tabular}{c||c|c}
\hline
$\ell$ &Here & DMRG\\
\hline
1 &0.49999999999999&0.5\\
2 &0.17659666969479&0.17659666969468\\
3 &0.04110985506014&0.04110985506012\\
4 &0.00595577151455&0.00595577151455\\
5 &0.00050690054232&0.00050690054232\\
6 &0.00002367077112&0.00002367077112\\
7 &0.00000055351689&0.00000055351689\\
\hline
\end{tabular}
\label{efp}
\caption{Emptiness formation probability of a chain of $20$ spins in the 
ground state for $\Delta=0.5$ (on the left). 
We compare our result (left) with those (numerically exact) from DMRG in Ref.
\cite{ccn-p}.}
\end{table}

There is another class of elements of $\rho_\ell$ accessible without 
further manipulations.  
In Eq. (\ref{action}) the term that is easily manipulated is
\begin{equation}
\pl_{b=1\atop{b\in\mathbf{A}_{j+1}}}^{M+j}\sinh(\la_b-\la_{a'_j})\,,
\label{denom}
\end{equation}
thus the only class with no poles is when in the sets 
$\mathbf{A}_j\,\mathbf{\alpha}_j\dots$ we have
\begin{equation}
\alpha^+=\{1\}\,,
\end{equation}
corresponding to the elements $P^{\e'_1,\dots,\e'_m}_{1,0,\dots,0}$, 
which is one particular column of the reduced density matrix.
For the other $2^\ell-1$ columns we need still further manipulations.

\subsection{A last trick for the general case}

The problems in the general case arise from the divergencies of the 
term (\ref{denom}).
Let us start with some preliminary observations.
First, it is important to know the maximum degree of the poles 
in (\ref{denom}).
Given a term of the summation in (\ref{action}), the order of the pole is the 
order of the zero in
\begin{equation}
\pl_{j\in \alpha^+}\pl_{b=M+1\atop{b\in\mathbf{A}_{j+1}}}^{M+j}\sinh(\la_{b}-\la_{a'_j})\,,
\end{equation}
that is given by
\begin{equation}
\sum_{i=1}^{\# a'_j>M}(\alpha^+_i-i)\,,
\end{equation}
where $\#a'_j$ is the number of elements $j\in \alpha^+$ such that $a'_j>M$.
It is easy to maximize the last expression to find
\begin{equation}
\sum_{i=1}^{d^+}{{}^{'}}(\alpha^+_{d^+-i+1}-i)\,,
\end{equation}
where the prime means that the summation is restricted to the $i$ such 
that $\alpha_{d^+-i+1}^+ -i>0$.

We construct the general procedure to tackle the poles in (\ref{denom}).
In the following we report our solution to the problem. 
However before proceeding we stress that this solution is 
somehow unsatisfactory, because we will end with and enormous sums of 
determinants for each element of $\rho_\ell$. 
If we would have been able to find a ``shortest'' representation of the 
same elements, we could have been able to describe much larger $\ell$.
Further developments in this direction would allow this method to be
competitive even with DMRG \cite{dmrg} for the ground-state.

Since the determinant in front of the pole is in general finite and the 
final result must be finite, all the coefficients multiplying each 
pole must sum to zero in (\ref{action}). 
Furthermore this implies that we can ignore these terms (because we know in 
advance that they give zero) and concentrate on the important ones.
To proceed, it is necessary to reshuffle the various terms in 
(\ref{action}). Let us define
\begin{equation}
\hat{G}=\frac{\pl_{j=1}^M \pl_{i=1}^M \sinh(\lambda_i -\mu_j +\eta)}
{\pl_{j>k}
\sinh(\mu_k-\mu_j)
\pl_{\alpha<\beta} \sinh(\lambda_{\b}-\lambda_{\a})}G\,,
\end{equation}
and
\begin{equation}
\hat{T}=\det T\,.
\end{equation}
We know that $\hat{T}\sim\e^{n(n-1)/{2}}$ in the homogeneous limit.
However here we have in general a pole of order $n(n-1)/{2}+\textrm{q}$ 
in $\hat{G}$, then we have to expand both $\hat{T}$ and the nonsingular part 
of $\hat{G}$  up to the order $n(n-1)/{2}+\textrm{q}$.
For $\hat{T}$, we developed the following procedure. 
Instead of doing the substitution (\ref{sub}), we put the higher orders in $\epsilon$
up to $n+{\rm q}$. Doing so we know that the determinant gives a polynomial in $\e$ with
lowest degree is $m_d=n(n - 1)/2$, and we indicate with $M_d$ the highest degree. 
%Doing so we know that the determinant
%gives a polynomial in $\e$ with lowest degree $n(n-1)/{2}$, while
%the maximum order in $\e$  is $(n+\textrm{q})n+n(n-1)/{2}$.
Thus the determinant is a polynomial of the form
\begin{equation}
\hat{T}=a_{m_d}\e^{m_d}+\dots+a_{M_d}\e^{M_d}
\end{equation}
%where we denoted with $m_d$ (minimum degree) the lowest order 
%$n(n-1)/{2}$ and 
%with $M_d$ (maximum degree) $n(n-1)/{2}+\textrm{q}$.
We can calculate all the coefficients $a_i$ numerically: 
it is enough to calculate the determinant in 
$D=M_d -m_d$ different points and then to invert the linear system.
Moreover if we choose the points in a smart way
\begin{equation}
p_k=\exp\bigg(\frac{2\pi i}{D}k\bigg)
\end{equation}
%where $D=M_d -m_d$, 
the solution of the linear system is numerically trivial since the matrix of the system is unitary.
Using this procedure we are able to calculate, in principle, all the elements 
of the reduced density matrix.
In practice, our possibilities are limited by the size of the density matrix.
Actually for small sizes we can go quite far and obtain the $\rho_\ell$ 
with three spins for chains with $200$ spins, a task impossible with exact 
diagonalization.

In Table \ref{tr3} we show the quantity $\Tr\rho_3^n$ for
$n=2$ and $n=3$ for odd chains at $\Delta=0.5$ where we can compare 
with the exact results in Ref. \cite{ncc-08}. 
The agreement is perfect and the small differences are due to the numerical 
rounding-off (we are summing order of $10^8$ elements in double precision
$10^{-16}$).

\begin{table}[t]
\centering
\begin{tabular}{c||c|c||c|c}
\hline
&
\multicolumn{2}{|c||}{${\rm Tr} \rho_3^2$}&
\multicolumn{2}{c}{${\rm Tr} \rho_3^3$}\\
\hline
$N$ &Here & Exact \cite{ncc-08}&Here & Exact \cite{ncc-08}\\
\hline
27  &    0.4130835714633  &  0.4130835714633&
0.1879727171090  &  0.1879727171090\\
51  &    0.4108297243638  &  0.4108297243637&
0.1851632322689  &  0.1851632322688\\
101 &    0.4101798729742  &  0.4101798729745&
 0.1843536264631  &  0.1843536264633\\
151 &    0.4100571750358  &  0.4100571750361&
0.1842007880727  &  0.1842007880729\\
201 &    0.4100139161598  &  0.4100139161591&
0.1841469044722  &  0.1841469044717\\
\hline
\end{tabular}
\label{tr3}
\caption{$\Tr\rho_3^2$ (left) and $\Tr\rho_3^3$ (right) for the ground-state 
of $\Delta=0.5$ obtained here compared with the known exact results.}
\end{table}

\subsection{Results: Entanglement entropy of excited states}
\label{Secres}

The main advantage of the method we have developed in the previous subsection
is that we can exactly evaluate all elements of the reduced density matrix 
for any eigenstate of the XXZ chain.
Compared to exact diagonalization we do not need to fully diagonalize the 
$2^N\times 2^N$ matrix to find the eigenstates, we can just pick up our 
desired state by choosing the correct quantum numbers. 
As we have already pointed out, once the eigenstate has been chosen, the 
numerical complexity of the algorithm is only a power-law in $N$ (actually $M$,
but for the most interesting states they are proportional), but the 
exponent grows linearly in $\ell$ limiting the range of applicability of 
the method. 
If we would have been interested only in the ground-state properties, this 
method is less effective than DMRG or any method based on matrix
product states \cite{dmrg}. 
In fact, these numerical methods require very little numerical 
effort to get the spectrum of the $\rho_\ell$ at machine precision
for the system sizes that are accessible to us. 
However it is hard, if not impossible, to calculate the entanglement properties
of highly excited states with DMRG. Thus our method, based on algebraic Bethe 
Ansatz, is by far the most effective available.
We checked that our algorithm reproduces the known results for the ground-state 
for several different $\Delta$, but we do not find instructive to report 
these results here.

For the study of excited states, we consider spin-chains of length $N=24$ 
in the critical antiferromagnetic region ($0<\Delta\le1$) for four different 
values of $\Delta=10^{-5},0.1,0.3,0.5$. 
Using our algorithm we generate the full reduced density matrices 
with $\ell\leq 6$ spins and from this we calculate the entanglement entropy 
(for the ground-state we know that already these small values of 
$\ell$ capture the asymptotic behavior \cite{ncc-08,ccn-p}).

\begin{figure}[t]
\includegraphics[width=0.49\textwidth]{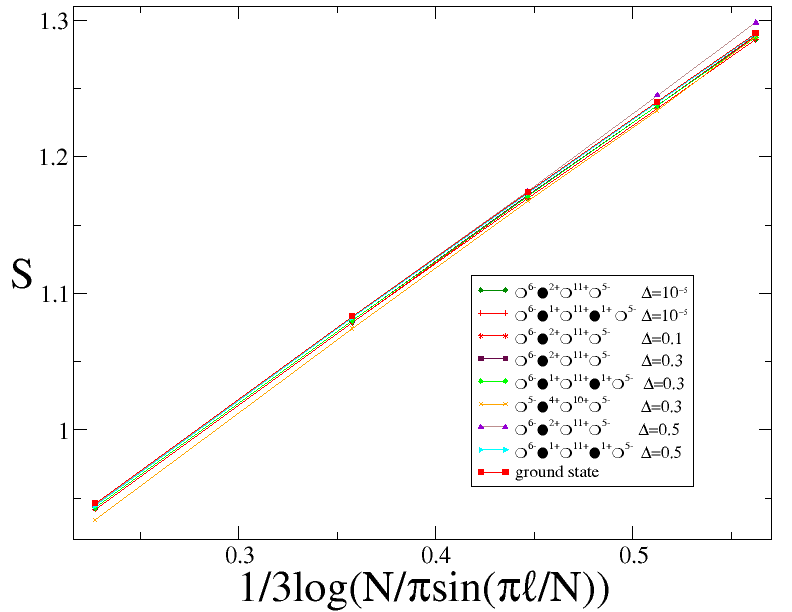}
\includegraphics[width=0.49\textwidth]{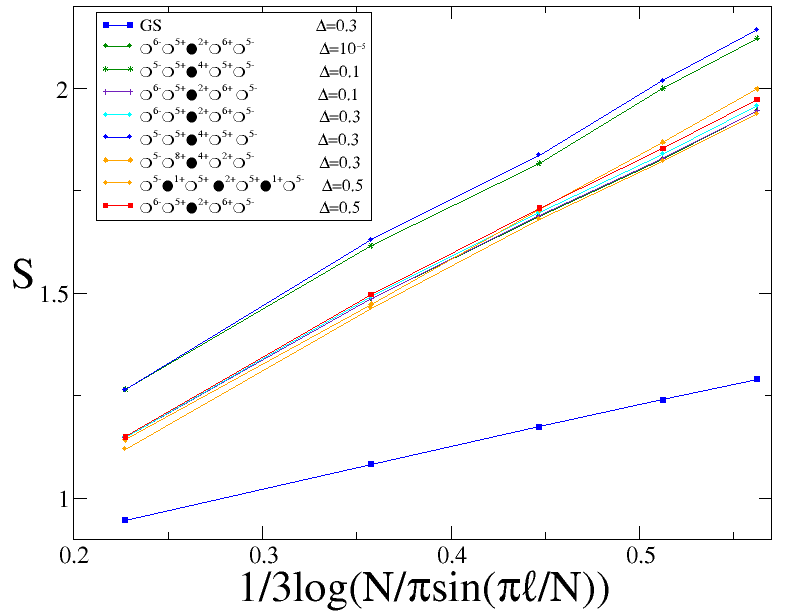}
\caption{
Entanglement entropy of the excited states of the XXZ spin-chain 
for $\Delta=10^{-5},0.1,0.3,0.5$
with $N=24$ plotted against the logarithm of the conformal distance.
Left: States that in the fermionic description for $\Delta=0$ have  
two discontinuities. 
The slope agrees with effective central charge equal $1$.
Right: States that in the fermionic description have four discontinuities. 
The results are compatible with an effective central charge equal to 2.
The bottom blue-line is the entropy of the ground-state at $\Delta=0.3$
shown for comparison.}
\label{fig2}
\end{figure}

\subsubsection{Bethe equations and classification of the states.}
The Bethe equations (\ref{bethe}) can be re-casted in a form that is useful for 
numerical solutions and for a complete classification of the states. 
For practical reasons, we consider only  the number of sites $N$ to be even.
We recall that the Hilbert space separates in sector with defined number of 
reversed spins $M$ (with respect to the reference state cf. Eq. (\ref{refstate})), 
that gives the total spin of the state in the $z$ direction $S_z^{TOT}=N/2-M$.
Taking the logarithm of Eq. (\ref{bethe}) and posing  $\zeta=\arccos(\Delta)$,
we have
\bea
\mbox{atan} \left[ \frac{\tanh(\lambda_j)}{\tan(\zeta/2)} \right] - 
\frac{1}{N} \!\sum_{k = 1}^M 
\mbox{atan} \left[\frac{\tanh(\lambda_j - \lambda_k)}{\tan \zeta}\right] =
\pi \frac{I_j}{N}\,.
\label{BAE_log}
\eea
Each set of {\it distinct} half-odd integer (integer) for $M$ even (odd) numbers 
$\{I_i\}$ (defined mod$(N)$)
specifies a set of rapidities, and therefore an eigenstate.
For example, in the ground state these numbers take the values
\begin{equation}
I_j^{(0)} = -\frac{M+1}{2} + j\,,\quad j = 1, ...,M\,.
\end{equation}
This ground state can be interpreted as the {\it spinon} vacuum. Spinons are the 
elementary excitations of the model. They have spin $1/2$ and obey 
semionic exclusion statistics (see e.g. \cite{kII} for a simple introduction to these
excitations).
Excited states have a defined number of up-spinon $n_+$ and down-spinon $n_-$.
The total number of spinons $n_++n_- \leq N$ is even when $N$ is even, while
$n_+-n_-= N-2M= 2 S^{TOT}_z$. (Actually for the interacting model 
different values of $S^z_{TOT}$ are possible when the state has 
some higher strings, which are non-dispersive. This discussion is too technical for the goals of 
this manuscript and we remand the interested reader to Ref. \cite{js-09}).
However, we will see that the spinon content is not the most important quantity for the 
entanglement entropy of excited states.

Employing the property that the quantum numbers $I_j$ are defined mod$(N)$,  we can choose 
the allowed ones in the sets
\be\fl
\left\{
\begin{array}{ll}\displaystyle
I^{\rm (odd)}=\{-\frac{N}{2},\dots,\frac{N}{2}-1\}      & 
{\rm for}\; M  %S_{TOT}^z =\frac{n_+ -n_- }{2} 
\; \textrm{odd}\,,\\ \\ \displaystyle
I^{\rm (even)}=\{-\frac{N}{2}+\frac{1}{2},\dots,\frac{N}{2}-\frac{1}{2}\}        
&  {\rm for}\;  M %S_{TOT}^z%=\frac{n_+ -n_- }{2} 
\; \textrm{even}\,.
\end{array}
\right.
\ee
Only a subset of these numbers, bounded by a calculable $I_{max}$ function of $\Delta$ and $M$ 
(see again Ref. \cite{js-09} for the technical details) provides real solutions for the rapidities 
$\lambda$, and we limited our attention to these states. 
Fixed the parity of $M$ (i.e. of $S_{TOT}^z$, since $N$ is even), any state is defined
by taking $M$ numbers among the allowed ones in $I^{\rm (odd)}$ or  $I^{\rm (even)}$.
The spinon content of the state then follows (see again \cite{kII}). 
Instead of using this standard $I_j$ notation to indicate the states,
following Ref. \cite{akmw-06}, we adopt a more complicated one 
that is useful to recover the fermionic description of the XX 
model when $\Delta=0$ (because we want to compare with the results in the 
previous section).
We denote with $\newmoon^{+(-)}$ the spinons with polarization up (down) 
and with $\fullmoon^{+(-)}$ the empty positions that can be occupied by 
spinons with up (down) polarization.
We indicate with an exponent the number of consecutive symbols, for instance
$\newmoon^{2+}\fullmoon^{11+}$ stands for 
\begin{equation}
\nm^+\nm^+\fm^{\,+}\fm^{\,+}\fm^{\,+}\fm^{\,+}\fm^{\,+}
\fm^{\,+}\fm^{\,+}\fm^{\,+}\fm^{\,+}\fm^{\,+}\fm^{\,+}\,.
\end{equation}
To each sequence of $N$ of these symbols $\nm^\pm$ and $\fm^\pm$ 
we can associate a single configuration of $I_i$.
For example, the sequence $\fm^{11-}\nm^{2+}\fm^{11+}$ has two up-spinons and no 
down ones; it follows that we need $M=N/2-(n^+-n^-)=11$ quantum numbers;
the state is fixed by taking from the set $I^{odd}$ the last $11$ numbers
(see again Ref. \cite{akmw-06} for more details).
The advantage of this maybe not really intuitive notation is that the rule to 
recover the fermionic description is very  easy: 
given the sequence one has to associate a fermion for 
every $\nm^-$ or $\fm^+$ \cite{akmw-06}.

Once we have the set $I_i$ for each state we are interested in, we use the 
Newton method to solve the Bethe equations.
We limit ourself to the two-spinon and four-spinon sector of the spectrum
(but we could easily consider other states). 
We also select states with real rapidities, to avoid problems with strings 
contributions, that however can be handled following Ref. \cite{js-09}. 

\subsubsection{Results.} 
The main feature we want to check here is if the conformal scaling 
(\ref{SAlog}) with an effective central charge $a$ is still valid for given 
excited states when we add the interaction $\Delta$ to the XX chain 
considered in the previous section. 
The prediction for the XX is based on the discontinuities of $\tilde m(\phi)$ (cf. 
Eq. (\ref{eq:mtilde})). 
In order to predict the result at $\Delta\neq 0$ we exploit the mapping
between the fermionic description and the spinonic one at $\Delta=0$. 
Once we have the fermionic picture associated to the state, we 
have $\tilde m(\phi)$ for $\Delta=0$. %so we can predict the behavior of the entropy.  
In order to check if the logarithmic scaling is obeyed, we plot $S_\ell$ 
against $S_\ell^{GS}$ in Eq. (\ref{resc}), 
so that if the dependence is linear, the slope gives automatically the
central charge $a$ of the effective Hamiltonian.
In Fig. \ref{fig2} (left) we display some states in the two-spinon 
and four-spinon sectors.
We choose these states in such a way that in the limit $\Delta\to0$,
the corresponding fermionic structure has two discontinuities. 
For example, the state $\fm^{6-}\nm^{2+}\fm^{11+}\fm^{5-}$ corresponds to
the fermion representation $|\downarrow^8\uparrow^{11}\downarrow^5\rangle$,
having two discontinuities in $\tilde m (\varphi)$.
For $\Delta=0$, we know from the previous section, that 
all these states are described by Eq. (\ref{SAlog})
with effective central charge $a=1$, as in the ground-state.
Fig. \ref{fig2} (left) provides a clear evidence that the asymptotic behavior 
of the entropy for $\ell\gg1$ does not depend on $\Delta$, at least in the 
considered range $\Delta\in [0,0.5]$. 
In the figure we also report the ground-state value for $\Delta=0.3$
for comparison.

\begin{figure}[t]
\includegraphics[width=0.49\textwidth]{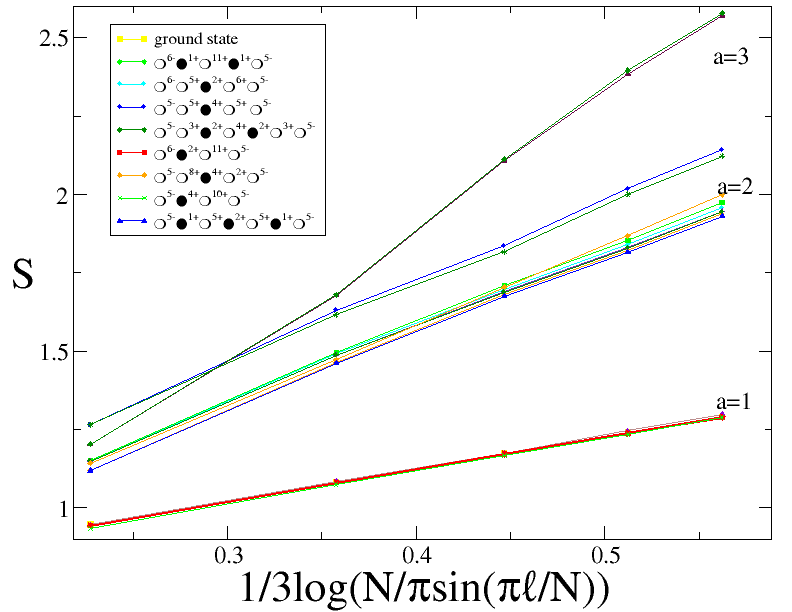}
\includegraphics[width=0.49\textwidth]{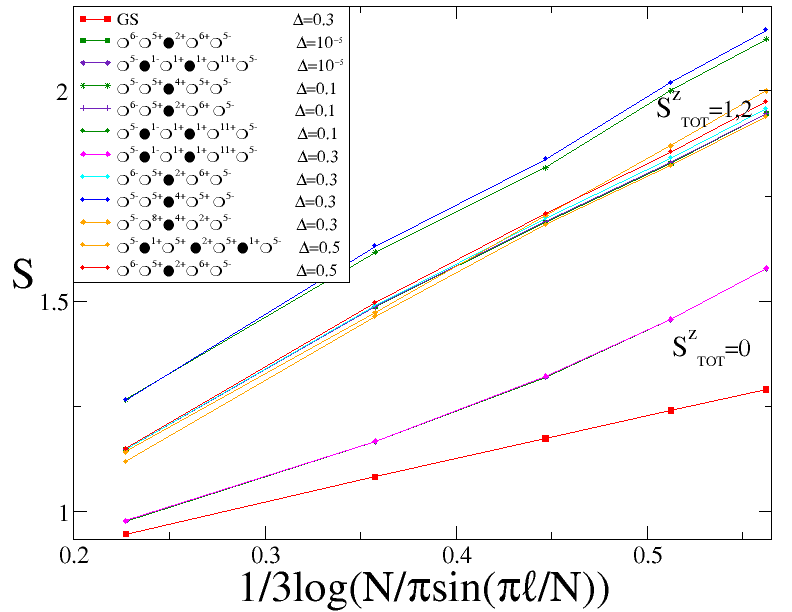}
\caption{Entanglement entropy for the two- and four-spinon states 
with $\Delta=10^{-5},0.1,0.3,0.5$.
Left: Summary of all the states we considered (for space problems, 
the legend shows only states at $\Delta=0.3$).
Right: Independence of the leading term on the spinon polarization.  
We considered $S_{TOT}^z=0,1,2$. 
The slope does not depend on the polarization.
The bottom-red line is the ground-state at $\Delta=0.3$. 
}
\label{fig3}
\end{figure}

In  Fig. \ref{fig2} (right) we report the entropy for some states whose 
fermionic description contains four discontinuities. 
Again we can observe that the data support the logarithmic behavior 
in Eq. (\ref{SAlog}). 
The slope is different from the ground-state one, and 
indeed a na\"{\i}ve fit (i.e. ignoring further corrections to the scaling that 
at $\ell\leq 6$ are important) of the constant $a$ gives $a\sim 2.3$,
which is in agreement with the XX prediction $a=2$. 
Moreover, the state $\fm^{5-}\fm^{5+}\nm^{4+}\fm^{5+}\fm^{5-}$
shows that the additive constant $c'_1$ in Eq. (\ref{SAlog}) depends 
dramatically on the details of the state (as we already know in the XX model).
In Fig. \ref{fig3} (right) we show the dependence on the 
spinon contribution of the additive constant. 
In Fig. \ref{fig3} (left) we report the von Neumann entropy for all states and 
values of $\Delta$ we calculated. 
The changing in behavior for different numbers of discontinuities is clearly 
visible. 
In this figure we also report two (almost indistinguishable) states that have six discontinuities
in the fermionic description and so are expected to have $a=3$. 
There are strong crossover effects preventing us to extract clearly the value
of $a$ for such small subsystems, but the data are clearly in the 
right direction. 
This crossover is expected from the results for the XX model: when having 
6 discontinuities in a chain of 24 spins, we expect approximately linear 
behavior in $\ell$ up to $\ell^*\sim N/6=4$, and in fact in the figure the crossover takes
place around $\ell\sim4$. A quantitative understanding of this crossover (even for more excited states) requires larger values of $\ell$ and $N$ that are not currently accessible to us.

\begin{figure}[t]
\includegraphics[width=0.65\textwidth]{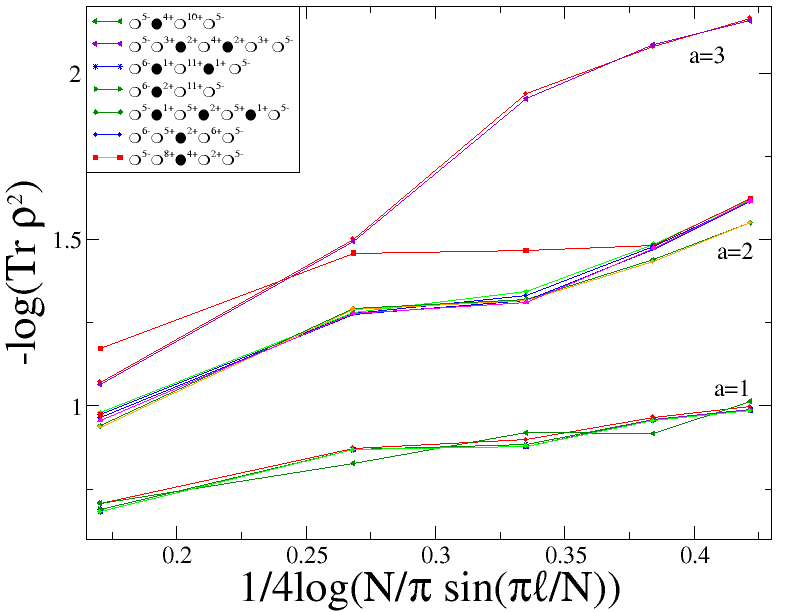}
\caption{$\log(\Tr\rho_\ell^2)$ against the logarithm of conformal distance.
In the legend we only give the states for $\Delta=0.3$.}
\label{fig4}
\end{figure}

To conclude this section, we also report in Fig. \ref{fig4} the data for 
$\log(\Tr\rho_\ell^2)$ plotted against the logarithm of the conformal distance 
to check the conformal prediction \cite{cc-04} 
\begin{equation}
-\log\Tr\rho_\ell^n=
\frac{1+n}{6n}c\log\bigg(\frac{N}{\pi}\sin\frac{\pi\ell}{N}\bigg)+c'_n\,,
\label{ren}
\end{equation}
for $n=2$. 
In fact, if the slope of all the previous curves can be interpreted as 
the central charge of some effective critical Hamiltonian having this state
as a ground-state, not only the entanglement entropy should follow
the conformal prediction (\ref{SAlog}), but also all R\'enyi entropies
should scale according to Eq. (\ref{ren}). 
And in fact, as for $S_\ell$ the curves arrange in sectors with approximately 
similar slopes. 
Strong even-odd oscillations of the R\'enyi entropies 
prevent us from any reliable quantitative analysis, as it is the case 
in the ground-state \cite{ncc-08}.
Again it is visible the same structure observed for the von Neumann entropy. 
However, na\"{\i}ve fits give reasonable estimations of the effective central 
charges for the two lowest sets, but the oscillations (combined with the
crossover previously mentioned) spoil the result for the last set for which 
$a=3$ is expect.

The knowledge of the full reduced density matrix can be also used to calculate the 
entanglement spectrum (i.e. the distribution of its eigenvalues). However, because of the 
relative small values of $\ell$ we can access, this is not enough to check recent 
conformal predictions for the spectrum \cite{cl-08}.

\section{Summary and discussions}
\label{concl}

In this paper we considered the entanglement entropy of excited states in spin chains.
We provided a full analytical study of the XY model in a transverse magnetic field. 
We found that the entanglement properties of the excited states depend strongly 
on the distribution of excitations above the ground state. 
To characterize them in the thermodynamic limit, we introduced the regularized 
characteristic function of excitation $m(\varphi)\in [-1,1]$. 
The analytic properties of this function (or slight variation of it for critical systems)
completely determine the entanglement in the scaling limit $N\gg\ell\gg1$. 
When $m(\varphi)\neq \pm1$ in a set of non-vanishing measure, we have that the 
entanglement entropy is extensive in the subset length (i.e. proportional to $\ell$).
The analytic expression for such states is given by Eq. (\ref{eq:extensive}) as we 
proved by using the Sz\"ego lemma for block Toeplitz matrices.
Oppositely when $m(\varphi)=\pm1$ almost everywhere (as for the ground state) 
the entanglement entropy always follows the conformal scaling (\ref{SAlog}), even
for non-critical systems (except in the ground-state when the area law holds).
The pre-factor (that we indicate as $a$) is the central charge of a critical, local, 
translational invariant Hamiltonian, that we built explicitly. 
In the case of the XX model we proved this result rigorously via the Fisher-Hartwig 
conjecture. These logarithmic states have a finite-size scaling that is by construction 
the conformal one in Eq. (\ref{SAlog}). 
Oppositely the extensive states have very peculiar finite size scaling with slopes 
that changes according to the analytic properties of $m(\varphi)$. 
We have been able to connect these features to the (non-)locality properties
of an effective Hamiltonian that can be made local on a wrapped chain.

We also considered the XXZ spin chain, that is solvable by Bethe Ansatz. 
We used the algebraic construction to calculate exactly the reduced density 
matrix for finite chains with $\ell\leq6$. 
The method we developed is ideal to obtain the entanglement entropy of excited states.
In fact, while numerical methods based on MPS like DMRG \cite{dmrg}
are very effective for the 
ground-state, they usually work bad for highly excited ones.
Our method instead treats on the same foot any eigenstate, that is specified by 
the quantum numbers related to the spinonic content of the state. 
This method has the numerical advantage that its complexity increases only in a 
polynomial way with $N$ (while exact diagonalization is exponential). 
The drawback is that the complexity increases exponentially with $\ell$ and limited
our study to $\ell\leq 6$. 
We do not not know whether this is an intrinsic limit of the method, or if our 
representation of the reduced density matrix can be still drastically optimized
to make the procedure more effective. The trickiest point in our derivation was 
to obtain the homogenous limit from the results in Ref. \cite{kmt-99,kmt-00}. 
If we would have been able to find a more effective way to perform this limit,
the method we propose could have been as effective as DMRG.
However, even if we could study only subsystem with $\ell\leq6$, we have been 
able to conclude that the main results obtained analytically for the XX model 
(at $\Delta=0$) remain valid when interaction is turned on. 
We showed in fact (making the proper mapping between spinonic and fermionic 
excitation at $\Delta=0$) that all the states that are logarithmic for $\Delta= 0$
maintain this property with the same prefactor and with a non-universal additive 
constant that depends very smoothly on $\Delta$ (as for the ground-state \cite{ccn-p}).

After this study, the characterization of the asymptotic block entanglement of excited states 
in these two chains is at an advanced level. Few unsolved problems are still present, 
especially for the XXZ chain, as e.g. the understanding of the string-states and the quantitative 
description of the crossover between linear and logarithmic behavior. 
However, the main question that still remains open is how general are these results.
The fact that the division among extensive and logarithmic states is conserved when the 
interaction $\Delta$ is introduced, strongly suggests that this phenomenon
should be expected for any local spin-chain, with a prefactor that 
can be predicted after that the relevant excitations have been identified.
In fact, in the interacting system (especially for  $\Delta$ not small) the excited states
are complicated linear combinations of the free-particle ones, several degenerations are 
also removed by $\Delta$, and it is unlikely that such result is only a coincidence.
However, we do not have a general proof for this statement. 
It might be that for low-lying excited states the generalization of the conformal 
methods of Refs. \cite{cc-04,cc-rev} can give the logarithmic behavior of these states.
Anyway, we have shown here that this property is not limited to low-lying states and 
so a more general proof would be desirable. 
Another very interesting question would be to understand how these results are 
affected by quenched disorder. For the ground-state it is known that only the prefactor 
of the logarithm is changed \cite{dis}, but the excitations in these systems are so 
different that major qualitative changes can take place.
Finally, it should be possible to generalize the methods employed for ground-states
of free lattice models in higher dimensional systems (as those reviewed in 
Ref. \cite{e-rev}) to the excited states of the same models. Some results in this direction, 
relevant for the physics of black holes are reported in Ref. \cite{ds-3}.

\section*{Acknowledgments}
We are extremely grateful to Jean-Sebastian Caux for his interest in this 
project and for continuous fruitful discussions.  
We thank G. Sierra and M. Ibanez for sharing with us their unpublished results
and for useful discussions.
We thank F. Colomo and F. Franchini for discussions.
PC benefited of a travel grant from ESF (INSTANS activity).

\end{document}